\documentclass[tradiabstract]{aa}

\usepackage{epsfig}
\usepackage{graphicx}
\usepackage{textcomp}
\usepackage{amsmath}
\usepackage{amsfonts}
\usepackage{amssymb}
\usepackage{booktabs}

\begin{document}

\title{Optical constants of refractory oxides at high temperatures}
\subtitle{Mid-infrared properties of corundum, spinel and $\alpha$-quartz,\\
potential carriers of the 13\,$\mu$m feature}
\author{S. Zeidler\inst{1}, Th. Posch\inst{2}, H.Mutschke\inst{1}}

\authorrunning{Zeidler et al.}
\titlerunning{Optical constants of refractory oxides}

\institute{Astrophysikalisches Institut und Universit\"atssternwarte, 
Friedrich-Schiller-Universit\"at Jena,
Schillerg\"asschen 2-3, D-07745 Jena, Germany\\
\email{harald.mutschke@uni-jena.de, simon.zeidler@uni-jena.de}
\and
Institut f\"ur Astrophysik, Universit\"at Wien, T\"urkenschanzstra{\ss}e 17,
A-1180 Wien, Austria\\
\email{thomas.posch@univie.ac.at}}


\date{Received date; accepted date}

\abstract
{Many cosmic dust species, among them refractory oxides, form at temperatures higher than 300\,K. Nevertheless, most astrophysical studies are based on the room-temperature optical constants of solids, such as corundum and spinel. A more realistic approach is needed for these materials, especially in the context of modeling late-type stars.}
{We aimed at deriving sets of optical constants of selected, astrophysically relevant oxide dust species with high melting points.}
{A high-temperature, high-pressure cell and a Fourier-transform spectrometer were used to measure reflectance spectra of polished samples.
For corundum ($\alpha$-Al$_2$O$_3$), spinel (MgAl$_2$O$_4$), and $\alpha$-quartz (SiO$_2$), temperature-dependent optical constants were measured from 
300\,K up to more than 900\,K. Small particle spectra were also calculated from these data.}
{All three examined oxides show a significant temperature dependence of their mid-IR bands. For the case of corundum, we find that the 13\,$\mu$m emission feature -- seen in the IR spectra of many AGB stars -- can very well be 
assigned to this mineral species. The best fit of the feature is achieved with oblate corundum grains at (mean) temperatures around 550\,K.
Spinel remains a viable carrier of the 13\,$\mu$ feature as well, but
only for T$<$300\,K and nearly spherical grain shapes. Under such
circumstances, spinel grains may also account for the 31.8\,$\mu$m band
that is frequently seen in sources of the 13\,$\mu$m feature and which
has not yet been identified with certainty.}



\maketitle

\section{Introduction \label{sec:intro}}

Within the past two decades, astromineralogy -- the science of solids in space -- has progressed significantly. Satellites such as the Infrared Space Observatory (ISO), {\em Spitzer}, and {\em Herschel} have made it possible to identify emission and absorption bands of crystalline solids in comets, in protoplanetary disks, in the shells of evolved stars, in supernova remnants, and in other environments including quasars (cf.\ Molster, Waters \& Kemper 2010). At the same time, the theory of formation and evolution of minerals in accretion disks and stellar outflows has contributed a lot to our understanding of the gas-solid transition in the cosmos (e.g., Gail 2010). Systematic studies of analog materials in terrestrial laboratories made it possible to establish databases of optical constants of solids that are expected or known to form from gases with a solar composition (e.g.\ Henning et al.\ 1999, Henning 2010). These databases continue to be indispensable sources of information on the UV to far-infrared properties of minerals and amorphous solids.

The present paper deals with refractory oxides, i.e.\ oxide minerals with
high melting points.
More specifically, we discuss the mid-infrared spectra of corundum
($\alpha$-Al$_2$O$_3$), spinel (MgAl$_2$O$_4$), and $\alpha$-quartz 
($\alpha$-SiO$_2$)\footnote{Note that SiO$_2$ and its polymorphs are sometimes classified as silicates because of their structural similarity to 
tectosilicates (see e.g.\ Klein, Hurlbut \& Dana 1999). Nevertheless, we will follow the widespread classification of quartz as an oxide here, following Strunz \& Nickel (2001).} measured at high temperatures (up to 973\,K).
Corundum, spinel, and quartz have repeatedly been proposed as carriers of infrared (IR) bands seen in astronomical objects (see below and Sect.\ 5). Owing to the lack of more comprehensive data, room temperature dielectric functions, and/or powder transmission, spectra of these minerals have been used in most previous papers.
However, the formation of refractory oxides at high temperatures -- and the fact that these oxides will partly {\em radiate}\/ at high temperatures and leave their spectral fingerprints in their high temperature state -- led us to the conclusion that their dielectric functions need to be derived by {\em in situ-}\/ high-T-measurements. This term is used to denote measurements of materials in their hot state, which is not the same as {\em annealing experiments.} In the latter case, a sample is heated to a certain temperature for a while. It may also have undergone an irreversible structural transition, and is afterwards cooled down to room temperature, at which point the spectra are then measured 
(see, e.g., J\"ager et al.\ 2011 for this distinction).

Thus, what we present here -- and for the first time in a systematic way -- are {\em in situ}\/ high-T-measurements of the IR spectra of 
$\alpha$-Al$_2$O$_3$, MgAl$_2$O$_4$, and $\alpha$-SiO$_2$, taking the effects of anisotropy into account where necessa  ry. Results of
annealing experiments with spinel can be found in Fabian et al.\ (2001).
The optical constants derived in the present paper will be made
publicly available via:\\
\texttt{http://www.astro.uni-jena.de/Laboratory/}\\
\texttt{Database/databases.html.}

This paper is structured in the following way: Section 2 describes the experimental methods and the samples that have been used.
In Sect.\ 3, we derive temperature-dependent oscillator
parameters from our reflectance measurements.
Section 4 presents small-particle spectra for different grain shapes.
In the final Sect.\ 5, we discuss astrophysical applications of our data,
focusing on the 13\,$\mu$m emission feature in AGB stars.


\section{Experimental methods}

\subsection{The high-temperature, high-pressure cell}
\label{HTHP}

For all our measurements, we used an IR Fourier-transform-spectrometer (Bruker 113v) equipped with a water-cooled, high-temperature, high-pressure (HTHP) cell. With this cell (Specac P/N 585 0), it is possible to heat samples to temperatures up to 1073\,K and to take IR spectra at the same time. Because of its dimensions, we had to deploy the cell into an aluminum tank that is placed in one of the two sample chambers of the Bruker spectrometer. The tank separates the cell {environment, which} is flooded with argon gas at atmospheric pressure, from the vacuum {inside} the spectrometer. The IR beam still can be applied for measurements by using two opposite windows in the tank. We used both potassium bromide (KBr) and polyethylene (PE) windows for measurements in the mid-IR (3--25\,$\mu$m) and far-IR (25--50\,$\mu$m) wavelength ranges, respectively.

\begin{figure}
 \begin{center}
 \includegraphics[width=\linewidth]{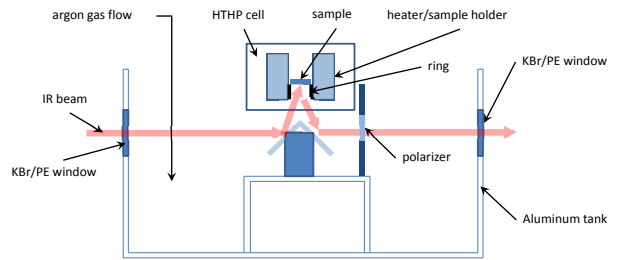}
\end{center}
\caption{Sketch of the set up of the HTHP cell in reflection mode in the aluminum tank.}
\label{HTHP-cell}
\end{figure}

The HTHP cell, set in reflectance mode, has been placed into the tank, where a base plate with two adjustable mirrors for the incoming and outgoing IR beam is used (see Fig.\,\ref{HTHP-cell}). The incident angle of the incoming beam on the surface of a sample in the cell is $\sim12^\circ$. We estimate the relative error in reflectance to the case of perpendicular irradiation to be $\approx3$\,\%. The cell itself has a closed sample chamber. The sample holder inside the cell is covered by the heater and has a diameter of 13\,mm and a length of 
$\sim$15\,mm. Special rings can be screwed into the holder to carry 
and/or to fix the sample.

The HTHP cell is supplied by an external temperature controller by which the heating temperature and the temperature gradient can be set. The controller measures the temperature of the sample holder and the body of the cell.
We additionally measured the temperature of the sample holder with an NiCr-Ni thermocouple. The temperatures, thus externally measured, are only by a few percentage points lower than the temperatures measured by the cell. All sample temperatures in this paper are externally measured ones.

\subsection{Samples and sample preparation}

Understanding the changes taking place in refractory oxides at high 
temperatures requires basic information on their internal structure. Therefore, we give crystallographic information on the respective solids at the beginning of each of the following sections, starting with Al$_2$O$_3$, the most refractory mineral treated here.

\subsubsection{Corundum}

The distribution of Al cations in either a face-centered cubic or a hexagonal close-packed lattice of oxygen anions results in a manifold of polymorphs of Al$_2$O$_3$. When heated, all polymorphs undergo phase transitions toward the thermodynamically most stable form, namely corundum (also $\alpha$-Al$_2$O$_3$;
Levin \& Brandon \cite{Levin98}). Corundum has a rhombohedral lattice structure. The oxygen anions are arranged in a (slightly distorted) hexagonal close-packing in which the aluminum cations occupy two thirds of the octahedral interstices. Due to the crystal structure, the interaction with light is not isotropic. The optical constants $n$ and $k$ (real and imaginary part of the complex refractive index) have to be calculated for polarizations perpendicular (ordinary) and parallel (extraordinary) to the crystal's symmetry (c) axis.

Corundum is a highly refractory material. {Its} condensation temperature is much higher than for most other minerals of astrophysical relevance. At pressures of $\sim$10$^{-9}$\,bar which occur in stellar outflows of evolved stars, corundum is already stable at more than 1400\,K, while silicates condense primarily below 1100\,K (Gail \cite{Gail10}). As a result, corundum should be one of the first condensates to form in the envelopes of M type stars. However, quantum mechanical calculations of the condensation rates have shown that corundum cannot form under these conditions by homogeneous condensation from the gas phase. Gail \& Sedlmayr (\cite{Gail98}) point out that the formation of Al$_2$O$_3$ grains is more likely to occur via grain-surface-reactions on already existing grain seeds of other high refractory materials such as TiO$_2$ or ZrO$_2$.

For the measurements discussed in this paper, we used a plane-parallel disk of pure synthetic corundum, which was manufactured by Thorlabs. This disk has polished surfaces and a diameter of 12.7\,mm, whereas the thickness is 3\,mm. Temperature-dependent reflectivity measurements on such pure corundum samples have been presented by Gervais \& Piriou (\cite{GP74}) for temperatures up to 1775\,K. In their paper, Gervais \& Piriou present a description of the four-parameter semi quantum (FPSQ) model and a comparison with the classical oscillator model. They show that the FPSQ model is a more effective way to fit the reflectivity of polar crystals in the case of wide reflection bands where equal damping of the longitudinal optical ($LO$) and transverse optical ($TO$) mode cannot be expected any more. On the other hand, Thomas et al.\ (\cite{Tho98}) developed a model that does not need a four-parameter description of the dielectric function but is based on 
adding temperature-dependent multiphonon contributions to a classical oscillator model. Thomas et al.\ list the parameters for the oscillators, together with their temperature dependence, corresponding to a second-order polynomial fit. They compared their model with measured data from several other authors and the FPSQ model. An application of the results of Thomas et al.\ to measured emission spectra of corundum in the wavelength range of 2.5-20\,$\mu$m is presented in the work of Sova at al.\ (\cite{Sova98}). However, the presented measurements on corundum in all of these papers refer only to the ordinary ray. To the authors' knowledge, high-temperature IR reflection spectra for the extraordinary ray that are necessary to calculate small particle spectra have not been measured yet. In this paper we present for the first time high-temperature IR reflection measurements of corundum for the ordinary and extraordinary ray and the calculate optical constants based on the FPSQ model. 

\subsubsection{Spinel}

In a wider sense, spinels are minerals with the general composition (A$_{1-x}$B$_x$)[A$_x$B$_{2-x}$]O$_4$. The atoms in parenthesis occupy the tetrahedral 
sites, and the atoms in square brackets occupy the octahedral sites of a cubic close-packed oxygen lattice.  The sum formula of pure Mg-Al-spinel is 
MgAl$_{2}$O$_{4}$. We will only refer to a material with this stoichiometry as spinel in the following.
Spinel crystals with this composition have an elementary cell consisting of 32\,oxygen anions, 16\,aluminum and 8\,magnesium cations. In a perfect spinel crystal, the trivalent Al cations are octahedrally coordinated by oxygen ions, while the bivalent Mg cations are tetrahedrally coordinated. However, it has been noted by Tropf \& Thomas (\cite{TroTho91}) and shown in detail by Fabian et al.\ (\cite{Fabian01}) that annealing induces the Mg and Al ions to change their sites, such that a part of the Al ions is occupying the tetrahedral sites and a part of the Mg ions is obversely located in the octahedral sites. This structural transition takes place at 1023--1073\,K and is apparently irreversible at least at short time scales (Tropf \& Thomas \cite{TroTho91}, Fabian et al.\ \cite{Fabian01}). Therefore, synthetic spinel crystals are in particular characterized by this deviation from the perfect crystal, since high temperature conditions are needed for their production. In contrast, natural spinel crystals that were exposed to low-temperature conditions for long periods show an almost perfect spinel crystal structure. Apparently, under low temperatures and after long periods, the aluminum cations leave the occupied tetrahedral sites and move to the octahedral sites, while the opposite is true for the magnesium cations (Fabian et al.\ \cite{Fabian01}). However, in the limiting case with zero Mg content, three eighths of the Al cations occupy tetrahedral sites, and the spinel transforms into $\gamma$-Al$_2$O$_3$.

Spinel has a very high melting point of $\sim$2300\,K at atmospheric pressure 
(Tropf \& Thomas \cite{TroTho91}) -- even higher than for terrestrial corundum.
 However, at the pressures prevailing in the dust-formation zones of AGB-stars
(in the range of $\sim$10$^{-8}$\,bar), corundum is able to form first
condensates at higher temperatures (at $\sim$1400\,K) than spinel
(which we expect to condense below 1150\,K). As mentioned above, spinel has a cubic crystal structure and is therefore optically isotropic. In our measurements we used a (pure) synthetic spinel disk (manufactured by SurfaceNet) with a diameter of 12.7\,mm and a thickness of 0.1\,mm. Like the corundum disk the spinel sample has polished surfaces. High-temperature data have already been obtained by Thomas et al.\ (\cite{Tho88}) and by Sova et al.\ (\cite{Sova98}). Thomas et al.\ published absorption coefficients in the range of 2.5-10\,$\mu$m for various temperatures up to 2000\,K and a comparison of the results with the multiphonon model. Sova et al.\ measured the emissivity of a pure spinel disk over a wavelength range of 2-12.5\,$\mu$m at $\sim$2000K and compared their results with the combined model of multiphonon contributions to the classical dielectric function from Thomas et al.\ (\cite{Tho98}). Our data cover the whole wavelength range from 5\,$\mu$m to 50\,$\mu$m.


\subsubsection{Quartz}

Quartz is one of the three low-pressure phases of crystalline SiO$_2$. The other two are tridymite and cristobalite.\footnote{The high-pressure phases of SiO$_2$ are coesite, stishovite, and seifertite. In total there are ten SiO$_2$ polymorphs (Klein, Hurlbut \& Dana 1999; El Goresy at al.\ 2008).}
It can be divided into a lower temperature, lower symmetry ($\alpha$) and a higher temperature, higher symmetry ($\beta$) modification. 

The mineral $\alpha$-quartz has a trigonal lattice structure at which the unit cell includes three silicon and six oxygen ions. The silicon cations are tetrahedrally coordinated by four oxygen anions which creates the [SiO$_4$]$^{4-}$-tetrahedron, the basic structure of almost all silicates and silica polymorphs. Each oxygen anion belongs to two tetrahedra, which leads to a three-dimensional grid of interconnected tetrahedra, with triples of them winding about a hypothetical axis that is parallel to the c-axis. Thereby, a helix-like structure is formed. It should be mentioned that these structures have no particular influence on the chemical and physical behavior of the crystal, they are just a help in visualizing the basic structure of $\alpha$-quartz.

The change between $\alpha$- and $\beta$-quartz is taking place at 846-847\,K at atmospheric pressure by a displacive phase transition (Lakshtanov et al \cite{Lak07}). $\beta$-quartz has a hexagonal lattice structure and its tetrahedra are arrangeded almost in the same way as for $\alpha$-quartz. The only difference lies in the alignment of the tetrahedra. In this respect, $\beta$-quartz has a higher symmetry than $\alpha$-quartz. By cooling down a $\beta$-quartz crystal, the opposite transition occurs, and an $\alpha$-quartz is formed again.

Quartz has been the topic of many studies, since it is one of the most important minerals on Earth and, with its physical and chemical properties, an outstanding material in science and technological applications. Spitzer \& Kleinman (\cite{SpKlein61}) were the first to examine the IR reflection properties of $\alpha$-quartz with respect to classical dispersion theory. They presented very comprehensive room-temperature studies of natural and synthetic quartz crystals. Gervais \& Piriou (\cite{GP75}) presented the first temperature-dependent IR reflection data of quartz. They fit their data between 295 and 975\,K with the already mentioned FPSQ model (Gervais \& Piriou \cite{GP74}) and analyzed the changes in the band parameters with temperature, especially in the range of the phase transition from $\alpha-$ to $\beta$-quartz.

\begin{figure}
 \begin{center}
 \includegraphics[width=\linewidth]{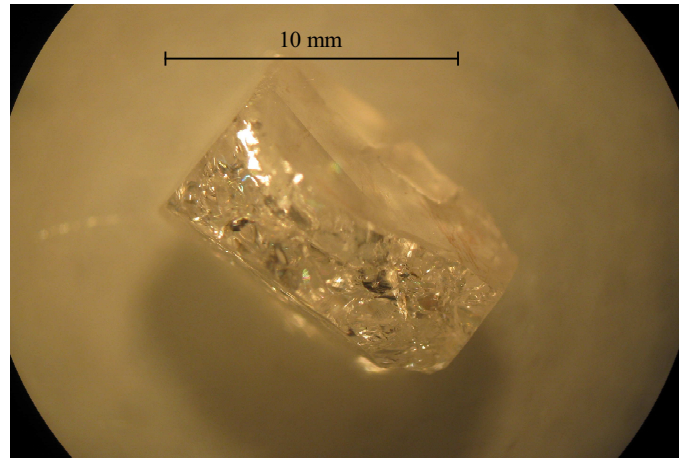}
\end{center}
\caption{Photograph of the quartz crystal that has been used for the measurements. The diameter of the field of view is 25\,mm.}
\label{quartz}
\end{figure}

In the present study, we took IR reflection measurements on a natural 
$\alpha$-quartz crystal from Brazil (see Fig.\ \ref{quartz}). The crystal has a maximum length, height, and width of 11, 7.5, and 8\,mm, respectively. EDX analysis did not reveal any case of impurities or deviations from stoichiometry. Because of the crystal structure of $\alpha$-quartz, the interaction with light is anisotropic. As in the case of corundum, one has to distinguish between polarizations parallel and perpendicular to the c-axis. Fortunately, because of the distinct hexagonal shape of the natural crystal used, we could locate the direction of the crystallographic c-axis quite easily. Moreover, the crystallographic c-axis lies parallel to the large natural grown surfaces of our $\alpha$-quartz crystal. Thus, spectroscopic measurements parallel and perpendicular to the c-axis could be taken on one of these surfaces. The crystal was cut (10\,mm $\times$ 5\,mm) so that it fit into the HTHP cell where the surface used for the measurements has been polished with diamond paste of 0.25\,$\mu$m fineness.

\subsection{Reflectance measurements}

We performed IR measurements with the HTHP cell in reflection mode 
(see Sect.\ \ref{HTHP}) in a wavelength range of 5-50\,$\mu$m. As mentioned above, special rings were used to hold the samples. There are two kinds of these rings: one with a hole of 12\,mm that was used to hold the synthetic samples and one with a hole of 5\,mm for the quartz crystal. For corundum and quartz, a polarizer was placed in the outgoing beam to distinguish between the different crystal orientations. We used two kinds of polarizers: a KRS-5 polarizer for mid-IR measurements (5-25\,$\mu$m) and a PE polarizer for far-IR measurements (25-50\,$\mu$m). With the aid of a low-resolution, real-time spectra shooting in the setup mode of the spectrometer software, the correct polarizer adjustment was established by finding the maximum of bands or by the disappearance of other bands.

As reference for each measurement, a spectrum of a gold mirror with a diameter of 12.7\,mm, taken at room temperature (300\,K), was used. Additionally, IR reflection spectra were taken after the measurements from the empty cell only equipped with the ring that was used for the respective sample. These empty measurements were also taken at the respective temperature of the measured sample and were necessary to exclude influences of the whole sample holder 
setup. The spectra of sample $R_{S}$ and empty cell $R_{E}$, both related to the reference, define the pure reflection spectra ($R$) via

\begin{align}
 R=\frac{R_{S}-R_{E}}{1-R_{E}}\,.
\end{align}

It could be verified that the heated samples and the sample holders themselves serve as additional IR sources whose emissions contribute to the spectra. While the radiation directly emitted toward the detectors gives a DC signal and is filtered out, the radiation back to the interferometer of the spectrometer becomes modulated and is superimposed on the modulated IR beam from the globar source of the spectrometer. Therefore, all interferograms would actually be superpositions of two sources. To get rid of the additional influence, we
measured the interferogram of the heated cell alone by decoupling the globar source and subtracted the result from each interferogram with activated globar source. After this, the resulting interferogram is Fourier-transformed into the desired (single channel) reflection spectrum that is used to calculate $R_S$ or $R_E$.

For the heating process, we used a constant temperature gradient of 10\,K/min to reach the respective measurement temperature. For corundum and spinel, we took mid-IR spectra at 300\,K, 551\,K, 738\,K, and 928\,K. For quartz, we also measured at 833\,K to derive more information on the development of the spectrum in the vicinity of the phase transition from $\alpha-$ to 
$\beta$-quartz. We chose no more than these individual temperatures 
to prevent the heater from stronger abrasion.


\section{Derived data and data analysis}\label{deriveddata}

\subsection{Derivation of the optical constants}

To calculate the optical constants from reflection measurements, we fitted our data with the FPSQ model of the dielectric function. This model provides the opportunity to fit measured data with independent {\em damping parameters}\/ for the $LO$ and the $TO$ mode, while the classical oscillator model only allows to assign different {\em frequencies}\/ to the above mentioned two modes. The necessity of assuming independent damping parameters arises in crystals with a strong mode splitting, where $LO$ and $TO$ phonon modes should have different phonon decay times (Gervais \& Piriou \cite{GP74}).

Based on the classical dispersion theory, the dielectric function can be written as:

\begin{equation}
 \epsilon=\epsilon'+i\epsilon''=\epsilon_{\infty}+\sum_j\Delta\epsilon_j\frac{\Omega^2_j}{\Omega^2_j-\omega^2-i\gamma_j\omega},
\label{eq:Lorentz}
\end{equation}

\noindent{}where $\Delta\epsilon_j$, $\Omega_j$, and $\gamma_j$ are the strength, resonance frequency, and damping of the $j$th oscillator. According to 
Maxwell's equations, the transverse modes are the poles of the dielectric function, and in contrast the longitudinal modes are the zeros. Barker (\cite{Barker64}) has shown that by multiplying all resonant denominators to Equation (\ref{eq:Lorentz}), a polynomial of degree 2$n$ is created ($n$ is the number of oscillators) and that this polynomial can be factorized with the zeros of the dielectric function. Consequently, the dielectric function is extended with the $LO$ frequency and damping parameters, and it can be rewritten in its alternative form:

\begin{equation}
 \epsilon=\epsilon_{\infty}\prod_j\frac{\Omega^2_{jLO}-\omega^2-i\gamma_{jLO}\omega}{\Omega^2_{jTO}-\omega^2-i\gamma_{jTO}\omega} .
\label{eq.:FPSQ}
\end{equation}

This is the FPSQ model in its simplest representation. A deeper treatment of this problem leads to the quantum form of the dielectric function (Gervais \& Piriou \cite{GP74}) that goes beyond the classical model toward a quantum mechanical description of the dispersion theory of anharmonic crystals. In our relatively simple case of fitting reflection data at high temperatures, the quantum form will not be important, though it becomes important in discussing the temperature dependence of the mode frequencies and the dampings, so it is rudimentarily discussed in Sect.\ \ref{sec:tempdep}.

The optical constants $n$ and $k$ (the real and the imaginary part of the refraction index) are a function of $\epsilon$ represented by the formulae:

\begin{equation}
 n=\sqrt{\frac{\epsilon'}{2}+\frac{1}{2}\sqrt{\epsilon'^2+\epsilon''^2}}
\label{eq.:n}
\end{equation}

\begin{equation}
 k=\frac{\epsilon''}{2n} .
\label{eq.:k}
\end{equation}

\noindent{}The reflection $R$ on a surface at near normal incidence in a vacuum can be expressed as a function of $n$ and $k$:

\begin{equation}
 R=\frac{(1-n)^2+k^2}{(1+n)^2+k^2} .
\end{equation}

\noindent{}Together with Eqs.\ (\ref{eq.:FPSQ}), (\ref{eq.:n}), and (\ref{eq.:k}) measured IR reflection spectra can be fit with the five parameters of this formula.

For all data, the fit procedure has been performed by using the {\it datafit}-function of the software SCILAB 5.4.0 (SCILAB consortium \cite{scilab}).


\subsubsection{Crystalline $\alpha$-Al$_2$O$_3$ (corundum)}

\begin{figure}
 \begin{center}
 \includegraphics[width=\linewidth]{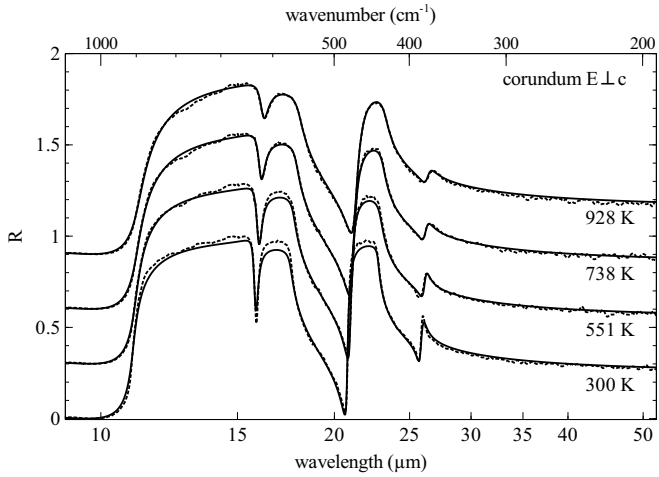}
\end{center}
\caption{Comparison of the fittings and the reflection spectra for the ordinary ray ($E_u$-type modes) of the corundum sample at four different temperatures. The fittings have been done with four oscillators.}
\label{corundum_refl_Esenk}
\end{figure}

\begin{figure}
 \begin{center}
\includegraphics[width=\linewidth]{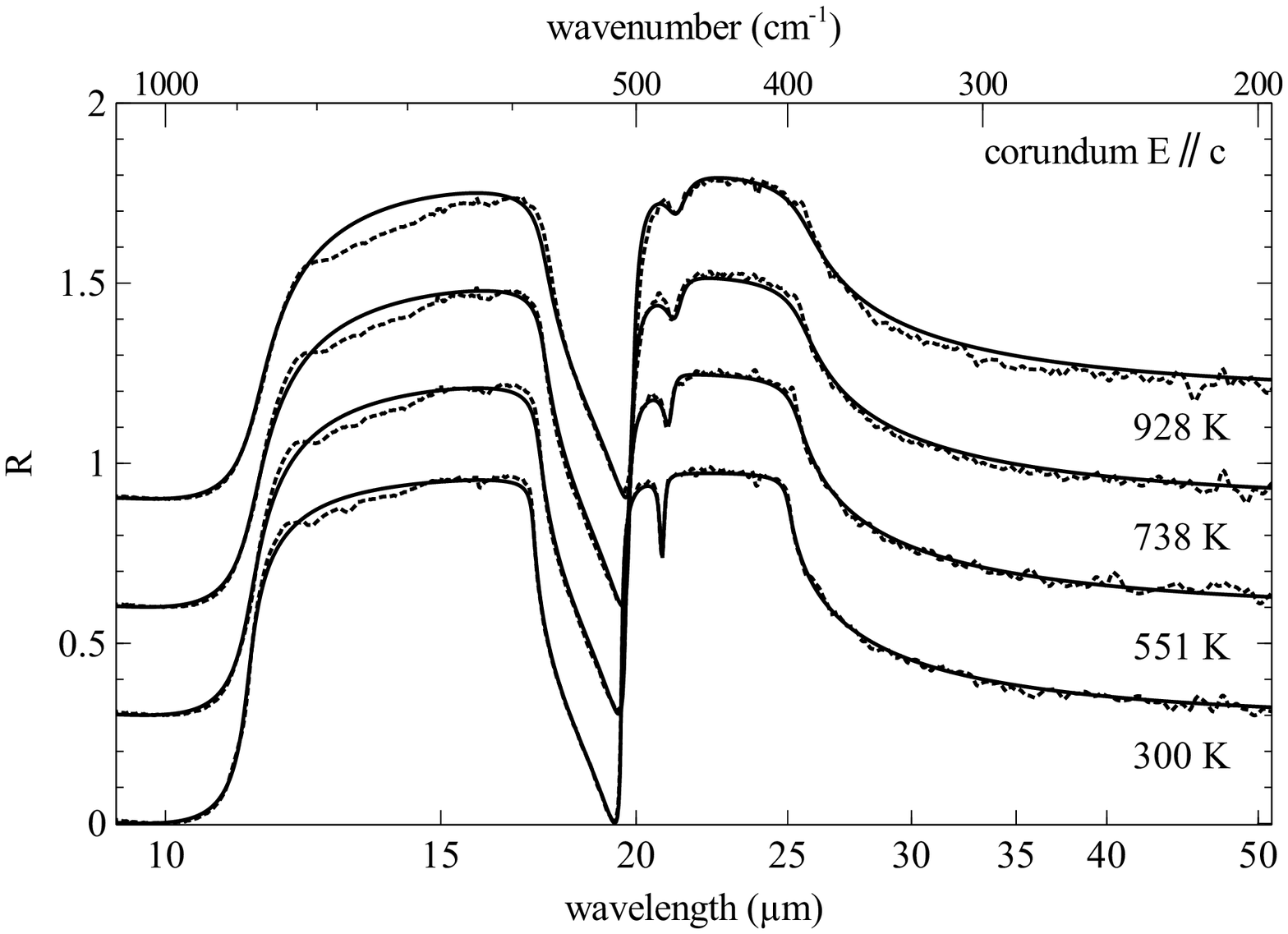}
\end{center}
\caption{Comparison of the fittings and the reflection spectra for the extraordinary ray ($A_{2u}$-type modes) of the corundum sample at four different temperatures. The fittings have been done with three oscillators.}
\label{corundum_refl_Epara}
\end{figure}

In Figs.\ \ref{corundum_refl_Esenk} and \ref{corundum_refl_Epara} the measured reflectivities of the corundum sample for the ordinary and the extraordinary ray at four different temperatures are shown, together with their best fit, with the aid of the FPSQ-model (Eq.\ (\ref{eq.:FPSQ})). As can be seen, the main changes due to heating occur in the band positions and the intensity of the bands. In general, the bands tend to shift toward longer wavelengths, while the intensity decreases with increasing temperature. This effect is due to the anharmonicity of the atomic potentials and can be seen more clearly in the values of the oscillator parameters listed in Tables \ref{tab:corfit} and \ref{tab:corfit2}. Here, the mode positions move toward longer wavelengths (lower wavenumbers, respectively), and the respective damping constants increase with increasing temperature.

According to the prediction of the irreducible representation of the vibration modes of corundum (Iishi \cite{Iishi78}), the spectra for the ordinary ray have been fit with four oscillators ($E_u$-type modes). In contrast, spectra for the extraordinary ray have been fitted with three oscillators, although theory predicts only two oscillators to be IR active ($A_{2u}$-type modes). Though, the third oscillator is needed to fit the distinct sharp minimum in the reflection spectra around 21\,$\mu$m. An additional band in that range might indicate
effects of surface polishing or of imperfect polarization.

The quality of the fits is different, depending on temperature and wavelength range. In general, the noise and the influence of water vapor in the FIR is much stronger than in the MIR, hence the fitting performance, especially for weak bands, is lower, and the quality of the resulting fit decreases for higher temperatures. As can be also seen in Figs.\ \ref{corundum_refl_Esenk} and \ref{corundum_refl_Epara}, the maxima of the measured reflectance spectra were not fit exactly. The measured values are either somewhat too high (for $E_u$-type modes) or too low (for $A_{2u}$-type modes). Barker (\cite{Barker64}) has shown that the kind of surface treatment can have a strong influence on the measured reflectance spectra profiles owing to additionally excited forbidden modes. Especially the structures on top of the strongest bands measured in both polarization directions give rise to the statement that the differences between measurement and fit in that range are caused by these forbidden modes. Likewise, these structures become weaker with increasing temperature.

\begin{table}
	\centering
	\caption{Comparison of the fitting parameters of corundum for the $E_u$-type modes (E$\perp$c) at different temperatures. The definition of the parameters refers to Eq.(\ref{eq.:FPSQ}).}
		\begin{tabular}{lllllll}
\toprule
 $j$ & $T$(K) & $\Omega_{TOj}$ &$\gamma_{TOj}$ & $\Omega_{LOj}$ & $\gamma_{LOj}$ & $\epsilon_{\infty}$ \\
 & & ($cm^{-1}$) & ($cm^{-1}$) & ($cm^{-1}$) & ($cm^{-1}$) & \\
\midrule
1 & 300 & 384.02 & 6.03 & 387.46 & 5.18 & 3.05\\
 & 551 & 380.83 & 6.13 & 383.22 & 6.55 & 3.06\\
 & 738 & 379.9 & 9.86 & 382.36 & 8.56 & 3.08\\
 & 928 & 377.51 & 10.97 & 379.8 & 10.1 & 3.1\\
\\
2 & 300 & 439.22 & 3.23 & 481.96 & 2.96 & -\\
 & 551 & 436 & 7.51 & 477.49 & 3.86 &\\
 & 738 & 432.94 & 8.19 & 474.39 & 6.99 &\\
 & 928 & 430.21 & 10.49 & 471.18 & 10.15 &\\
\\
3 & 300 & 569.35 & 7.86 & 908.23 & 22.37 & -\\
 & 551 & 564.42 & 8.89& 901.64 & 27.84 &\\
 & 738 & 560.1 & 10.24 & 898.5 & 36.78 &\\
 & 928 & 555.46 & 14.82 & 894.5 & 45.36 &\\
\\
4 & 300 & 634.36 & 5.6 & 630.59 & 8.53 & -\\
 & 551 & 628.8 & 8.13 & 624.95 & 11.37 &\\
 & 738 & 624.51 & 10.8 & 620.9 & 14.32 &\\
 & 928 & 619.99 & 14.21 & 616.66 & 17.55 &\\
\bottomrule
\end{tabular}
	\label{tab:corfit}
\end{table}

\begin{table}
	\centering
	\caption{Comparison of the fitting parameters of corundum for the $A_{2u}$-type modes (E$\parallel$c) at different temperatures ($j=1,2$). The third oscillator could not be verified as an $A_{2u}$-type mode.}
		\begin{tabular}{lllllll}
\toprule
$j$ & $T$(K) & $\Omega_{TOj}$ &$\gamma_{TOj}$ & $\Omega_{LOj}$ & $\gamma_{LOj}$ & $\epsilon_{\infty}$\\
 & & ($cm^{-1}$) & ($cm^{-1}$) & ($cm^{-1}$) & ($cm^{-1}$) & \\
\midrule
1 & 300 & 399.68 & 4.68 & 511.05 & 1.42 & 2.9\\
 & 551 & 394.76 & 11.2 & 508.09 & 2.79 & 2.91\\
 & 738 & 391.91 & 18.58 & 505.24 & 2.88 & 2.91\\
 & 928 & 390.54 & 22.57 & 502.65 & 3.22 & 2.97\\
\\
2 & 300 & 481.58 & 3.42 & 480.93 & 3.21 & -\\
 & 551 & 476.38 & 7.83 & 475.7 & 6.99 &\\
 & 738 & 472 & 13.3 & 471 & 12.13 &\\
 & 928 & 471.19 & 17.11 & 469.84 & 16.32 &\\
\\
3 & 300 & 582 & 4.17 & 884.75 & 21.57 & -\\
 & 551 & 577.36 & 8.93 & 882.32 & 35.96 &\\
 & 738 & 573.5 & 11.37 & 878.47 & 44.71 &\\
 & 928 & 573.35 & 15.74 & 874.39 & 49.24 &\\
\bottomrule
\end{tabular}
	\label{tab:corfit2}
\end{table}

 
\subsubsection{Crystalline MgAl$_2$O$_4$ (spinel)}

\begin{figure}
 \begin{center}
\includegraphics[width=\linewidth]{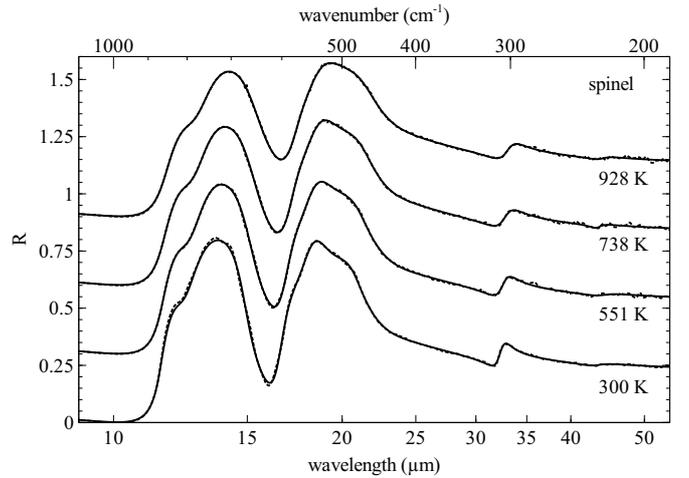}
\end{center}
\caption{Reflection spectra of the spinel crystal at four different temperatures.}
\label{spinel_refl}
\end{figure}

In Fig.\ \ref{spinel_refl} the spinel reflection spectra (dotted lines) are compared with their best fit (solid lines) at four different temperatures. The fits are in very good agreement with the measured data and can be treated as almost perfect models for the measurements within the accuracy given by the line thickness.

For a normal, non-defective MgAl$_2$O$_4$ crystal, group theory predicts only four IR active oscillators whose $TO$-modes are located at around 15, 18, 21, and 33\,$\mu$m (Thibaudeau \cite{Thi02}). However, the weak band at 43\,$\mu$m, as well as the weak shoulders at around 12 and 19\,$\mu$m, made it necessary to use eight oscillators for the fitting procedure (see Table \ref{tab:spfit}). These additional modes are forbidden by group theory but can be excited by the mentioned disordered cation distribution in synthetic spinels (Fabian et al.\ (\cite{Fabian01})) and by defects in the crystal lattice similar to the case of corundum (see above).

Previous studies have also made use of additional modes by fitting reflection spectra of spinel. Fabian et al.\ (\cite{Fabian01}) used a classical Lorentz oscillator model and derived fits with eight oscillators in a wavelength range of 9--35\,$\mu$m, which means they did not treat the 43\,$\mu$m band. Therefore, they used an extra mode at 11.4\,$\mu$m ($S$-mode) to fit a weak shoulder on the short-wavelength edge of the broad band between 11--15\,$\mu$m, which they link to the Al/Mg-ratio of their spinel samples and the partial filling of tetrahedral interstices by Al- instead of Mg-ions. In our spectra we could not see any trace of such a shoulder. Thibaudeau et al.\ (\cite{Thi02}) applied the FPSQ-model to their spectra with six oscillators. Based on their results, we added two more oscillators to the short- and the long-wavelength edges of the broad 19\,$\mu$m band to fit the shoulder at 17\,$\mu$m and the decrement toward longer wavelengths correctly. These extra oscillators have already been found in previous studies (Fabian et al.\ (\cite{Fabian01}), O'Horo et al.\ (\cite{OHoro72}), Preudhomme \& Tarte (\cite{Preud71})) and might be due to the ratio of Mg/Al-cation disorder and defects, respectively.

Comparing the results of the fitting parameters listed in Tab.\ \ref{tab:spfit}, it is noticeable that only the strongest oscillators show a monotone development of their parameters with the temperature. The weaker oscillators ($j=1,3,6$) do not exhibit such distinct behavior, which might be because in general the fitting of weak modes strongly depends on the quality of the measured spectra, making the fits non-unique (e.g.\ the weak band around 43$\mu$m). However, in general the mode positions move toward longer wavelengths upon heating, while their respective damping increases with temperature.

\begin{table}
	\centering
	\caption{Comparison of the fitting parameters of spinel. Oscillators 2,4,5, and 7 represent the predicted $T_{1u}$-type modes of the theory.}
		\begin{tabular}{lllllll}
\toprule
 $j$ & $T$(K) & $\Omega_{TOj}$ &$\gamma_{TOj}$ & $\Omega_{LOj}$ & $\gamma_{LOj}$ & $\epsilon_{\infty}$ \\
 & & ($cm^{-1}$) & ($cm^{-1}$) & ($cm^{-1}$) & ($cm^{-1}$) & \\
\midrule
1 & 300 & 231.05 & 8.97 & 231.58 & 8.29 & 2.77 \\
 & 551 & 226.96 & 9.99 & 227.54 & 9.41 & 2.8 \\
 & 738 & 230.02 & 9.79 & 230.34 & 9.18 & 2.82 \\
 & 928 & 227.33 & 13.81 & 227.78 & 13.45 & 2.77\\
\\
2 & 300 & 307.12 & 12.65 & 310.73 & 13.68 & -\\
 & 551 & 304.4 & 16.38 & 308.36 & 17.02 &\\
 & 738 & 302.2 & 20.15 & 306.49 & 20.68 &\\
 & 928 & 300.07 & 20.15 & 304.01 & 20.58 &\\
\\
3 & 300 & 427.11 & 134.26 & 442.85 & 112.26 & -\\
 & 551 & 426.34 & 139.54 & 438.44 & 108.65 &\\
 & 738 & 438.12 & 140.41 & 444.8 & 96.33 &\\
 & 928 & 440.49 & 140.09 & 442.56 & 97.23 &\\
\\
4 & 300 & 481.53 & 30.58 & 608.03 & 37.88 & -\\
 & 551 & 475.24 & 35.89 & 598.4 & 42.79 &\\
 & 738 & 469.66 & 38.7 & 590.55 & 48.62 &\\
 & 928 & 466.52 & 45.67 & 582.36 & 55.59 &\\
\\
5 & 300 & 531.37 & 55.1 & 558.16 & 70.14 & -\\
 & 551 & 527.69 & 63.05 & 553.85 & 74.68 &\\
 & 738 & 521.27 & 65.86 & 549.66 & 79.03 &\\
 & 928 & 520.32 & 76.53 & 546.93 & 82.07 &\\
\\
6 & 300 & 581.8 & 74.64 & 546.32 & 120.96 & -\\
 & 551 & 576.95 & 74.12 & 547.55 & 122.79 &\\
 & 738 & 569.98 & 75.11 & 542.25 & 126.64 &\\
 & 928 & 563.9 & 73.29 & 544.39 & 126 &\\
\\
7 & 300 & 672.28 & 36.67 & 866.53 & 39.52 & -\\
 & 551 & 662.28 & 47.55 & 862.97 & 47.55 &\\
 & 738 & 655.42 & 55.98 & 860.15 & 55.98 &\\
 & 928 & 649.32 & 64.24 & 855.98 & 64.24 &\\
\\
8 & 300 & 811.63 & 77.37 & 800.67 & 73.77 & -\\
 & 551 & 797.84 & 86.36 & 785.72 & 76.12 &\\
 & 738 & 789.47 & 92.02 & 776.78 & 79.87 &\\
 & 928 & 786.4 & 93.61 & 772.56 & 85.91 &\\
\bottomrule
\end{tabular}
	\label{tab:spfit}
\end{table}


\subsubsection{Crystalline $\alpha$-SiO$_2$ (quartz)}

\begin{figure}
 \begin{center}
 \includegraphics[width=\linewidth]{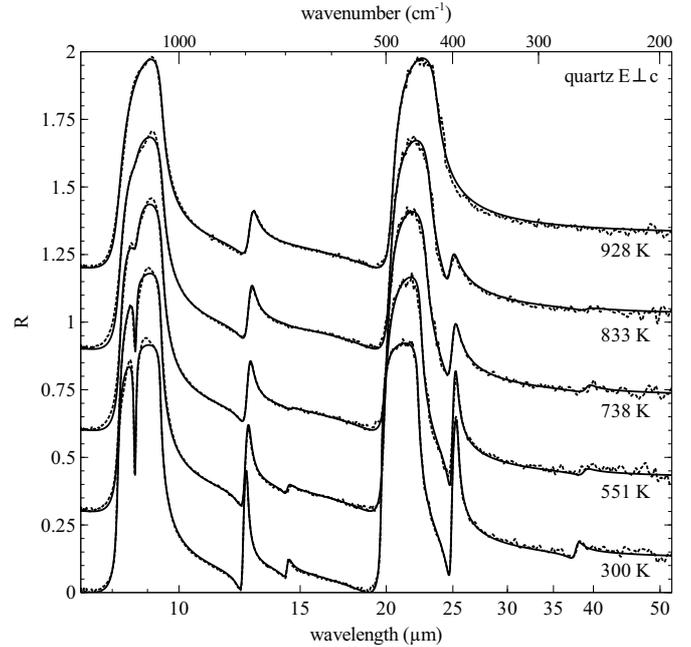}
\end{center}
\caption{Reflection spectra of the quartz crystal in the ordinary ray at five different temperatures ($E$-type modes).}
\label{quartz_refl_Esenk}
\end{figure}

\begin{figure}
 \begin{center}
 \includegraphics[width=\linewidth]{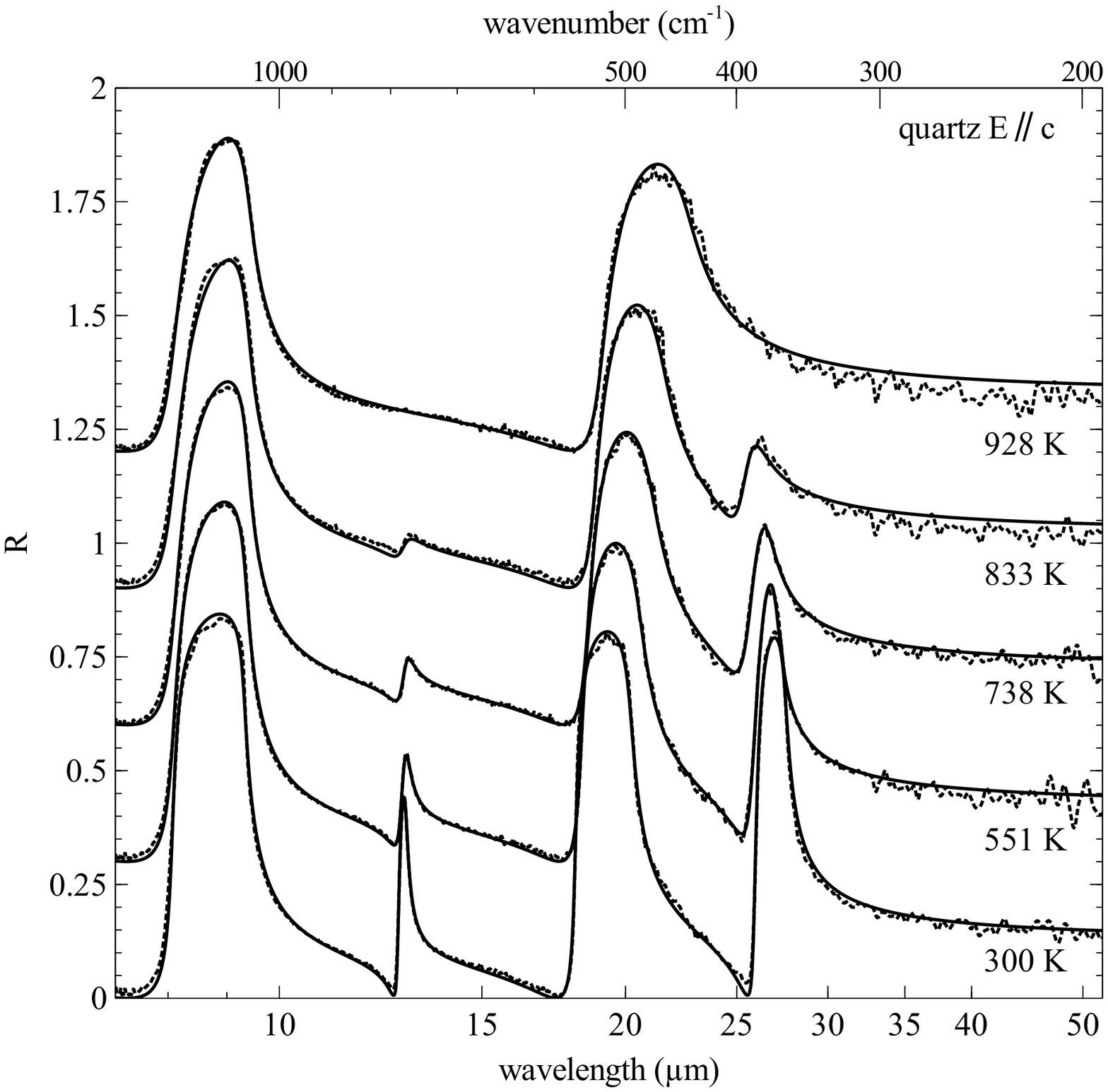}
\end{center}
\caption{Reflection spectra of the quartz crystal in the extraordinary ray at five different temperatures ($A_2$-type modes).}
\label{quartz_refl_Epara}
\end{figure}

In contrast to spinel and corundum, the spectra of quartz show a lot of distinct narrow bands, as can be seen in Figs.\ \ref{quartz_refl_Esenk} and \ref{quartz_refl_Epara} for the ordinary and the extraordinary rays, respectively. Theoretical calculations predict eight $E$- and four $A_2$-type modes to be IR active in $\alpha$-quartz, corresponding to the excitations in the ordinary and extraordinary rays, respectively. By contrast, in $\beta$-quartz, four $E$- and only two $A_2$-type modes should be IR active as analyzed by Scott and Porto (\cite{Scott67}). Two $A_2$ modes remain from the $\alpha$-$\beta$-transition of quartz, while the other two become forbidden. The same holds true for the $E$ modes, but here the missing modes only become IR inactive, while they still remain Raman active (Scott and Porto \cite{Scott67}).

The fitting parameters for the ordinary and the extraordinary ray at five different temperatures are shown in Tables \ref{tab:qufitEsenk} and \ref{tab:qufitEpara}. We found seven $E$- and four $A_2$-type modes for temperatures below 847\,K. The eighth $E$-type mode should be found at wavelengths around 78\,$\mu$m, which is, however, beyond our covered wavelength range and therefore has not been measured. Anyway, the parameter values are quite consistent with data given by the literature (Scott and Porto \cite{Scott67}, Gervais \& Piriou \cite{GP75}) and they fit our measured data very closely (see Figs.\ \ref{quartz_refl_Esenk} and \ref{quartz_refl_Epara}). 

By reaching 928\,K for the ordinary ray, it can be clearly seen that the bands at 8.2, 14, 25, and 37\,$\mu$m (mode number 7, 4, 2, and 1) disappeared. According to Scott and Porto, these are the four modes that should become IR inactive by the transition to $\beta$-quartz. Already at 833\,K, the modes 7, 4, and 1 became very weak so that there is almost no evidence of their existence in the spectrum. In the extraordinary ray, the spectra also show the transition to $\beta$-quartz. Here, modes number 1 and 3 (at 26 and 13\,$\mu$m) disappeared by reaching 928\,K, as predicted by theory. These results correspond very well to the data taken by Gervais \& Piriou (\cite{GP75}), who comprehensively investigated the phase transition of quartz with IR spectroscopy between 7.7 and 33.3\,$\mu$m. At this point we cab complement their study with data on the temperature development of the $E$-type mode at 37\,$\mu$m that has not been treated in their work. 

The quality of the fits decreases at longer wavelengths due to noise and absorption bands of water vapor in the measured spectra. In those with very strong water-absorption bands, we tried to keep the fits slightly above the measured spectra at longer wavelengths since the signal intensity had 
decreased in that range (see e.g.\ Fig.\ \ref{quartz_refl_Epara}).

\begin{table}
	\centering
	\caption{Comparison of the fitting parameters of quartz for the ordinary ray.}
		\begin{tabular}{lllllll}
\toprule
 $j$ & $T$(K) & $\Omega_{TOj}$ &$\gamma_{TOj}$ & $\Omega_{LOj}$ & $\gamma_{LOj}$ & $\epsilon_{\infty}$ \\
 & & ($cm^{-1}$) & ($cm^{-1}$) & ($cm^{-1}$) & ($cm^{-1}$) & \\
\midrule
1 & 300 & 264.13 & 6.68 & 265.36 & 7.08 & 2.34 \\
 & 551 & 257.69 & 9.21 & 258.36 & 9.05 & 2.36 \\
 & 738 & 255.06 & 9.29 & 255.77 & 9.87 & 2.38 \\
 & 833 & 256.4 & 11.18 & 256.69 & 10.94 & 2.39 \\
 & 928 & - & - & - & - & 2.36 \\
\\
2 & 300 & 393.55 & 4.25 & 402.88 & 4.33 & -\\
 & 551 & 395.03 & 5.26 & 401.96 & 4.88 &\\
 & 738 & 396.5 & 10.44 & 402.99 & 11.05 &\\
 & 833 & 400.14 & 12.7 & 404.6 & 11.83 & \\
\\
3 & 300 & 449.78 & 3.43 & 507.89 & 2.87 & -\\
 & 551 & 445.61 & 4.9 & 504.58 & 7.88 &\\
 & 738 & 438.27 & 9.28 & 499.84 & 9.29 &\\
 & 833 & 432.7 & 11.19 & 495.08 & 10.62 & \\
 & 928 & 423.04 & 11.85 & 491.45 & 12.18 &\\
\\
4 & 300 & 694.3 & 11.91 & 697.31 & 12.11 & -\\
 & 551 & 695.37 & 13.54 & 697.13 & 12.52 &\\
 & 738 & 688.59 & 15.17 & 688.99 & 14.51 &\\
 & 833 & 680.64 & 38.36 & 680.51 & 37.27 & \\
\\
5 & 300 & 795.36 & 7.75 & 808.72 & 8.11 & -\\
 & 551 & 790.2 & 13.44 & 804.58 & 13.02 &\\
 & 738 & 786.14 & 16.97 & 799.52 & 16.62 &\\
 & 833 & 782.98 & 19.92 & 796.81 & 20.2 & \\
 & 928 & 780.28 & 22.85 & 793.68 & 26.32 &\\
\\
6 & 300 & 1064.57 & 6.98 & 1226.34 & 10.91 & -\\
 & 551 & 1063.3 & 9.31 & 1221.61 & 18.66 &\\
 & 738 & 1058.36 & 15.17 & 1217.97 & 20.43 &\\
 & 833 & 1057.85 & 20.31 & 1220.69 & 31.12 & \\
 & 928 & 1060.37 & 19.53 & 1226.38 & 43.43 &\\
\\
7 & 300 &  1160.03 & 8.38 & 1157.02 & 7.98 & -\\
 & 551 & 1157.62 & 10.57 & 1155.85 & 9.98 &\\
 & 738 & 1155.9 & 36.83 & 1153.29 & 34.45 &\\
 & 833 & 1154.85 & 41.18 & 1154.19 & 39.93 & \\
\\
\bottomrule
\end{tabular}
	\label{tab:qufitEsenk}
\end{table}

\begin{table}
	\centering
	\caption{Comparison of the fitting parameters of quartz for the extraordinary ray.}
		\begin{tabular}{lllllll}
\toprule
 $j$ & $T$(K) & $\Omega_{TOj}$ &$\gamma_{TOj}$ & $\Omega_{LOj}$ & $\gamma_{LOj}$ & $\epsilon_{\infty}$ \\
 & & ($cm^{-1}$) & ($cm^{-1}$) & ($cm^{-1}$) & ($cm^{-1}$) & \\
\midrule
1 & 300 & 363.03 & 5.18 & 386.42 & 3.76 & 2.39 \\
 & 551 & 369.72 & 7.89 & 388.87 & 11.92 & 2.38 \\
 & 738 & 376.11 & 15.02 & 392.48 & 18.9 & 2.38\\
 & 833 & 386.59 & 20.1 & 396.64 & 20.05 & 2.38 \\
 & 928 & - & - & - & - & 2.44\\
\\
2 & 300 & 494.99 & 10 & 552.82 & 5.3 & -\\
 & 551 & 486.1 & 17.85 & 547.22 & 10.28 &\\
 & 738 & 473.93 & 23.83 & 540.15 & 14.12 &\\
 & 833 & 465.2 & 24.9 & 533.84 & 17.53 & \\
 & 928 & 442.2 & 30.04 & 527.43 & 24.19 & \\
\\
3 & 300 & 775.58 & 8.48 & 789.17 & 6.51 & -\\
 & 551 & 775.09 & 13.78 & 785.37 & 15.29 &\\
 & 738 & 775.27 & 19.98 & 782.47 & 20.58 &\\
 & 833 & 777.02 & 31.45 & 782.29 & 30.85 & \\
\\
4 & 300 & 1073.37 & 15.21 & 1238.42 & 16.41 & -\\
 & 551 & 1070.08 & 20.03 & 1237.02 & 30.04 &\\
 & 738 & 1066.91 & 22.76 & 1234.71 & 39.73 &\\
 & 833 & 1065.91 & 26.52 & 1239.85 & 50.93 &\\
 & 928 & 1064.14 & 32.9 & 1240.89 & 55.58 &\\
\\
\bottomrule
\end{tabular}
	\label{tab:qufitEpara}
\end{table}


\subsection{The temperature dependence of the oscillator 
parameters\label{sec:tempdep}}

\subsubsection{Theoretical point of view}

The lattice potential energy of a solid is, in general, anharmonic 
(Maradudin and Fein \cite{Mara62}, Cowley \cite{Cow63}). The 
anharmonicity has, in principle, two main effects on the solid and its spectral properties: thermal expansion of the solid and interactions of phonons of different modes influence their energy and give them a finite lifetime (the 
reciprocal counterpart to the damping). If the phonon interactions are sufficiently small, the phonon frequency $\Omega_j$ for the $j$th 
mode as introduced in (\ref{eq.:FPSQ}) can be expressed in terms of an unperturbed quasiharmonic mode frequency $\omega_j$ and a frequency shift $\Delta\omega_{PIj}$ representing the influence of the phonon interactions. According to Gervais and Piriou (\cite{GP74},b), $\Omega_j$ takes the form

\begin{equation}
\Omega_j^2=\omega_j^2+2\omega_j\Delta\omega_{PIj}.
\label{eq:Omega}
\end{equation}

Owing to the anharmonicity, both $\omega_j$ and $\Delta\omega_{PIj}$,\footnote{It should be noted that $\Delta\omega_{PIj}$ in general carries a 
frequency dependence that still can be neglected in the FPSQ model 
(Gervais and Piriou \cite{GP74}). Also, $\Omega_j$ is written here as 
representative of both $LO$ and $TO$ mode frequencies.} hence $\Omega_j$, show 
a temperature dependence. While for $\Delta\omega_{PIj}$ the temperature dependence arose from the mode-specific excitation probability of the phonons ({\it Bose-Einstein}-distribution), the temperature dependence of $\omega_j$ is due to the thermal expansion of the solid and can be approximated by the model of {\it Gr\"uneisen} (Lowndes 
\cite{Low70}, Jasperse et al.\ \cite{Jas66}, Gervais and Piriou \cite{GP74}) 
with:

\begin{equation}
\omega_j=\omega_{0j}+\delta\omega_j=\omega_{0j}-\omega_{0j}\int_0^T{g_j(T')\alpha(T')dT'},
\label{eq:omega}
\end{equation}

\noindent{}where $\omega_{0j}$ represents the phonon-mode frequency at 0\,K for the $j$th mode. Here, $g_j$ is the respective mode Gr\"uneisen parameter and 
$\alpha$ the volume thermal expansion of the material. While for most 
of the materials $\alpha$ can be approximated well by a linear 
temperature dependence, $g_j$ has a rather complex relation to the 
temperature (Jasperse et al.\ \cite{Jas66}), which is often unknown. 
However, in many cases $g_j$ can be assumed as a constant value ranging 
between 0 and 2 (Lowndes \cite{Low70}, Gervais and Piriou \cite{GP75}), 
in which case $\delta\omega_j$ will show a quadratic temperature dependence.

\begin{figure*}
  \begin{center}
  \includegraphics[width=0.75\linewidth]{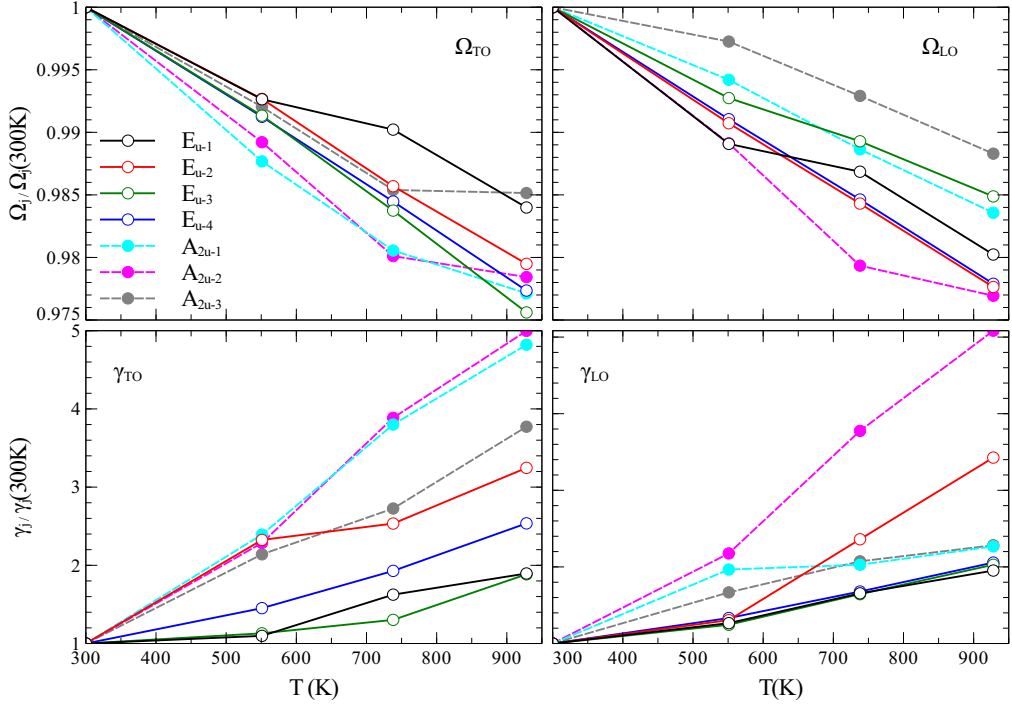}
\end{center}
\caption{Comparison of the temperature development of the mode 
parameters of corundum as given in Tables \ref{tab:corfit} and 
\ref{tab:corfit2}. The parameters have been normalized to their 
respective value at 300\,K and are ordered with respect to $TO$ and $LO$ 
mode, as well as to frequency and damping.}
\label{Tdep1}
\end{figure*}

The phonon-interaction contribution to the frequency shift 
$\Delta\omega_{PIj}$ has been determined by Maradudin and Fein 
(\cite{Mara62}) and by Cowley (\cite{Cow63}) involving three phonon processes to second-order and four phonon processes to first-order pertubation theory. With their assumptions, they obtain 
a linear temperature dependence of the frequency shift in the high-temperature limit. A quadratic temperature dependence can be obtained by 
including terms of higher order pertubation theory and/or higher phonon processes as shown by Ipatova 
et al.\ (\cite{Ip67}) and Gervais and Piriou (\cite{GP74-2}). By assuming 
$\Delta\omega_{PIj}$ to be sufficiently small, equation (\ref{eq:Omega}) 
can be simplified to

\begin{equation}
\Omega_j=\omega_{0j}+\delta\omega_j+\Delta\omega_{PIj}.
\label{eq:Omegasimp}
\end{equation}

Maradudin and Fein (\cite{Mara62}) pointed out that the frequency shift $\Delta\omega_{PIj}$ appears as the real part of a complex phonon self energy. In their theory, the imaginary part is represented by a function $\Gamma_j$ that has the property of altering the phonon energy, so it can be seen as a representative of the damping function. Like $\Delta\omega_{PIj}$, damping appears as a consequence of the phonon interaction due to anharmonicity, specifically as 
the inverse of a now finite phonon lifetime. The quantity $\Gamma_j$ is related to the classical damping factor $\gamma_j$ by the equation

\begin{equation}
\gamma_j=\Gamma_j\left(\frac{2\omega_{0j}}{\omega}\right)_{\omega=\Omega_j}.
\label{eq:Gamma}
\end{equation}

Like the frequency shift, the damping generally shows a quadratic 
temperature dependence in the high temperature limit (Ipatova et al.\ 
\cite{Ip67}, Gervais and Piriou \cite{GP74-2}).

\begin{table*}[htbp]
\centering
\caption{Comparison of the temperature-fit parameters of corundum 
for the modes in the ordinary ray ($E_u$-type modes).}
\begin{tabular}{lllll}
\toprule
$j$ & $\beta_{j1}$ & $\beta_{j2}$ & $\gamma_{j1}$ & $\gamma_{j2}$ \\
& ($10^{-3}cm^{-1}K^{-1}$) & ($10^{-6}cm^{-1}K^{-2}$) & 
($10^{-3}cm^{-1}K^{-1}$) & ($10^{-6}cm^{-1}K^{-2}$) \\
\midrule
1 ($TO$) & -9.84 & 0.2 & 0 & 6.6 \\
2 ($TO$) & -10.67 & -3.1 & 12.02 & 0 \\
3 ($TO$) & -14.2 & -6.5 & 0 & 7.9 \\
4 ($TO$) & -19.91 & -2.4 & 2.61 & 9 \\
1 ($LO$) & -13.62 & 1 & 3.38 & 3.7 \\
2 ($LO$) & -18.39 & 1 & 0 & 8.9 \\
3 ($LO$) & -23.54 & 1 & 0 & 30 \\
4 ($LO$) & -22.72 & 0.4 & 6.05 & 6.8 \\

\bottomrule
\end{tabular}
\label{tab:Tfit1}
\end{table*}

\begin{table*}[htbp]
     \centering
     \caption{Comparison of the temperature fit parameter of corundum 
for the modes in the extraordinary ray ($j=1,3$: $A_{2u}$-type modes).}
         \begin{tabular}{lllll}
\toprule
  $j$ & $\beta_{j1}$ & $\beta_{j2}$ & $\gamma_{j1}$ & $\gamma_{j2}$ \\
& ($10^{-3}cm^{-1}K^{-1}$) & ($10^{-6}cm^{-1}K^{-2}$) & 
($10^{-3}cm^{-1}K^{-1}$) & ($10^{-6}cm^{-1}K^{-2}$) \\
\midrule
1 ($TO$) & -17.12 & 1 & 29.21 & 0 \\
2 ($TO$) & -19.65 & 1 & 15.48 & 5.4 \\
3 ($TO$) & -17.02 & 1 & 15.53 & 2.1 \\
1 ($LO$) & -9.34 & -3.4 & 3.25 & 0 \\
2 ($LO$) & -20.59 & 1 & 8.25 & 10.5 \\
3 ($LO$) & 3.28 & -16.3 & 47.94 & 0 \\
\bottomrule
\end{tabular}
     \label{tab:Tfit2}
\end{table*}

\subsubsection{The case of corundum}

The temperature dependence of the complete set of mode 
parameters for corundum is shown in Fig.\,\ref{Tdep1}. For this comparison we 
chose to plot the relative changes of the parameters with respect to 
their respective values at 300\,K, similar to the low temperature 
forsterite data presented in Koike et al.\ (\cite{Koike06}). A first 
look shows that the thermal shift of the resonance frequencies 
$\Omega_{TO}$ (upper left panel) is relatively similar for all modes; 
i.e., the $\Omega_{TO}$ values change by 1.5\%-2.5\% for 900~K relative 
to 300~K, while for all other parameters describing damping and strength 
of the resonances the differences among the modes are considerably 
larger. For some reason, the damping parameters
of the $A_{2u}$-type modes increase more than for the $E_u$-type modes, 
however, no other clear trends (e.g.\ strong vs.\ weak modes) can be found.

A fitting of the temperature dependence with respect to the given theory 
is quite difficult because of the limited temperature range and 
resolution, especially without any data for lower temperatures. 
Nonetheless, we try fitting both $\Omega_j(T)$ and $\gamma_j(T)$ by the 
quadratic equations

\begin{equation}
\Omega_j(T)=\Omega_j(300)+\beta_{j1}(T-300)+\beta_{j2}(T^2-300^2)
\label{eq:fitFreq}
\end{equation}
and
\begin{equation}
\gamma_j(T)=\gamma_{j}(300)+\gamma_{j1}(T-300)+\gamma_{j2}(T^2-300^2),
\label{eq:fitDamp}
\end{equation}

\noindent{}respectively. In these, we have subtracted the values at 300\,K on both 
sides, leaving the coefficients $\beta_{j1}$, $\beta_{j2}$, 
$\gamma_{j1}$, and $\gamma_{j2}$ unchanged with respect to a 0\,K-based fitting curve. The coefficients have been

restricted such that $\Omega_j(T)$ is ensured to monotonically 
decrease, at least until the melting point of corundum 
($\approx$2300\,K), while $\gamma_j(T)$ has to  monotonically increase. 
Thus, $\beta_{j1}$ should be normally negative, and $\beta_{j2}$ is not 
allowed to be larger than $1\times10^{-6}$\,$cm^{-1}K^{-2}$, whereas 
$\gamma_{j1}$ and $\gamma_{j2}$ are both restricted to positive values.

The results are listed in the Tables \ref{tab:Tfit1} and 
\ref{tab:Tfit2}. As can be seen, only the last $A_{2u}$-type mode has a 
positive $\beta_{j1}$, which is however compensated for by a large negative 
$\beta_{j2}$, indicating that the shift of the mode becomes much stronger 
at high temperature. For some of the $E_u$ and most of the $A_{2u}$-type 
modes, $\beta_{j2}$ reaches the positive limit of 
$1\times10^{-6}$\,$cm^{-1}K^{-2}$, which indicates that the data imply a 
significant weakening of the temperature trend at high T. Generally, 
{the absolute value of} $\beta_{j1}$ tends to increase with increasing 
mode frequency, which is because we fit the absolute 
frequencies  here, having already noted above (for the $TO$ 
frequencies) that the behavior of the relative frequency changes would 
be similar among the modes.

The behavior of the $\gamma_j(T)$ with temperature appears to be quite 
diverse, as indicated by the fitting parameters. For the $A_{2u}$-type 
modes, where the damping is generally rising more sharply with the temperature 
than for the $E_u$-type modes, the changes can more easily be 
represented by a linear fitting (large $\gamma_{j1}$) and the trend 
tends to be flattened at higher T ({that would result in a negative 
$\gamma_{j2}$ but is ``zero'' due to our restriction}). In contrast, for 
the modes in the ordinary ray ($E_u$-type modes), considerable positive 
quadratric behavior ($\gamma_{j2}$ dominating) is found.

Trying to distinguish between thermal expansion effects and the phonon-interaction contribution in the frequency shift, we write $\Delta\omega_{PIj}=\Delta_{PIj1}T+\Delta_{PIj2}T^2$ and set $g_j$ as temperature-independent ({Eq}.\,\ref{eq:omega}). Then, $\beta_{j1}$ and $\beta_{j2}$ can be expressed by

\begin{equation}
\begin{split}
\beta_{j1}&=\Delta_{PIj1}-\omega_{0j}g_j\alpha_0\\
\beta_{j2}&=\Delta_{PIj2}-\omega_{0j}g_j\alpha_1/2,
\end{split}
\label{eq:alpha}
\end{equation}

\noindent{}where $\alpha_0$ and $\alpha_1$ are the coefficients of the linear thermal expansion $\alpha=\alpha_0+\alpha_1T$. For corundum, they have been determined by Fiquet et al.\ (\cite{Fi99}) to $\alpha_0=2.081\times10^{-5}$ and $\alpha_1=6.6\times10^{-9}$\,K$^{-1}$. At room temperature, Lodziana and Parlinski (\cite{Lod03}) give an interval for all mode Gr\"uneisen parameters of corundum ranging from 0.4 to 1.8 with an averaged value $\left\langle g\right\rangle=1.14$. They find that the highest values (1.8) are reached by the $A_{2u}$ and the lowest (0.4) by the $E_u$-type modes. We adopt these values for all temperatures.

Taking the room-temperature resonance frequencies as an estimate of the $\omega_{0j}$, we can derive, for the latter ($E_u$-type modes, $g_j=[0.4,1.14]$), a significant $\Delta_{PIj1}$ in the range between $-14.63\times10^{-3}$ and $-0.7\times10^{-3}$\,$cm^{-1}K^{-1}$. Similar to the mentioned development of $\beta_{j1}$, the absolute value of $\Delta_{PIj1}$ also tends to increase with increasing $\Omega_j(300)$. On the other hand, $\Delta_{PIj2}$ can be positive or negative depending on the mode, and it lies between $-5\times10^{-6}$ and $3.09\times10^{-6}$\,$cm^{-1}K^{-2}$. It shows no clear development with increasing $\Omega_j(300)$.

The estimated values of $\Delta_{PIj1}$ for the $A_{2u}$-type modes in the limits $g_j=[1.14,1.8]$ are only for $j=1$ in the same magnitude as the ones for the $E_u$-type modes. In the case of $j=3$, they vary between $-3.2\times10^{-3}$ and $4.8\times10^{-3}$\,$cm^{-1}K^{-1}$. Nevertheless, $\Delta_{PIj2}$ is positive for both modes, and it lies between $4\times10^{-6}$ and $7.91\times10^{-6}$\,$cm^{-1}K^{-2}$. We found an overall increase in the values of both $\Delta_{PIj1}$ and $\Delta_{PIj2}$ with increasing $\Omega_j(300)$. {Since both mechanisms (phonon interaction and thermal expansion) generally have negative contributions in their linear coefficients, the calculations confirm that the values of $\beta_{j1}$ are reasonable.}

\subsubsection{Spinel and quartz}

For spinel, the temperature fit is much more difficult due to the weak 
bands and the unknown degree of ion disorder that can have a big 
influence on the fits in the case of a synthetic spinel (Thibaudeau et 
al.\ \cite{Thi06}). Therefore, because of the missing agreement with the 
mode parameters, we will not show the temperature fits of spinel.

The temperature dependence of the phonon frequencies and dampings in 
quartz differs from the assumptions that have been made so far. Quartz already
undergos a phase transition at 847\,K that has a strong influence 
on the behavior of the bands. Gervais and Piriou (\cite{GP75}) found 
that in the vicinity of the phase transition the temperature dependence 
of the frequency shift for $\alpha$-quartz is almost only related to the 
thermal expansion. Furthermore, the dampings show a linear temperature 
dependence up to 700\,K as would appear if only cubic anharmonicity 
were taking place, while a critical increase appears in the approach of the 
phase transition. In contrast, all bands that survived the phase 
transition to $\beta$-quartz do not show a strong temperature behavior 
any longer, even in the vicinity of the transition temperature. This 
behavior is similar to that of the thermal expansion (Gervais and Piriou 
\cite{GP75}). Another interesting point is the temperature behavior of 
the frequencies of the modes $(E)_{j=2}$ and $(A_2)_{j=1}$. It shows a 
monotone increase in both $TO$ and $LO$ vibrations, whereas all other 
modes show a more ore less smooth decrease. Both bands disappear by 
reaching the transition point.

The possible temperature development of $\epsilon_{\infty}$ has been 
neglected in the FPSQ calculations of Gervais and Piriou (\cite{GP74} 
and \cite{GP75}) for corundum and quartz as well as in those of 
Thibaudeau et al.\ (\cite{Thi06}) for spinel. However, Cowley 
(\cite{Cow63}) has shown that $\epsilon_{\infty}$ has a temperature 
dependence when taking multiphonon contributions into account. 
Therefore, while fitting our data, we allowed $\epsilon_{\infty}$ to 
vary and could indeed verify a certain temperature dependence for all 
measured materials, with the general trend toward an increase in 
$\epsilon_{\infty}$ with increasing temperature. This result is 
supported by the data of Thomas et al.\ (\cite{Tho98}), who compared 
measured near IR refractive indexes of corundum with their multiphonon 
model. They found a linear increase of $n$ with increasing temperature, 
corresponding to a quadratic increase of $\epsilon_{\infty}$ 
(since the absorption is too small to count for $\epsilon$ in that 
wavelength range). Unfortunately, we cannot give any
analytical temperature dependence of $\epsilon_{\infty}$ 
until we have more data at higher and lower temperatures, 
which would increase the accuracy of the resulting fit.


\section{Calculation of small-particle spectra \label{sec:small particle}}

Using the dielectric function $\epsilon$ (see Eq.\ \ref{eq.:FPSQ})
that was calculated during the fitting procedure, we
derived the absorption cross section for spherical particles small compared to the wavelength with the aid of the theory by Gustav Mie (Mie \cite{Mie08}). 
If we let $a$ be the grain radius and $\lambda$ the wavelength, then

\begin{equation}
C_{abs}=\frac{8\pi^2a^3}{\lambda}{Im}\left(\frac{\epsilon-1}{\epsilon+2}\right).
\label{eq:Mie}
\end{equation}

For nonspherical particle shapes, a useful method for calculating 
small-particle spectra is the {\em distribution of form factors}\/ (DFF) model 
(Min \cite{Min06}). In the framework of this model, the absorption cross section is given by a generalization of Eq.\ (\ref{eq:Mie}) in the form

\begin{equation}
C_{abs}=\frac{8\pi^2a^3}{\lambda}\int_0^1Im\frac{\epsilon-1}{1+L(\epsilon-1)}\frac{P(L)}{3}dL,
\label{eq:DFF}
\end{equation}

\noindent{}where $P(L)$ is a distribution function of form factors $L$ that are 
defined in the interval (0;1) and contain the information about the particle shape. In this case, $a$ is the radius of a volume-equivalent sphere.
For $P(L)=\delta(L=1/3)$, Eq.\ (\ref{eq:Mie}) is obtained. Mutschke et al.
(\cite{Mutschke09}) have demonstrated that certain synthetic DFFs derived by Min et al.\ (\cite{Min06}) for aggregates of spherical particles and for Gaussian random sphere particles are well suited to reproducing measured 300\,K absorption spectra of typical corundum and spinel powders with either more roundish or more irregular grain shapes, respectively. We calculate such absorption spectra at high temperatures in addition to those of spherical particles to demonstrate the diversity of spectra that can be derived from 
our optical constants.

For materials with anisotropic optical behavior like quartz and 
corundum, the absorption cross section has to be calculated for each incoming-beam polarization parallel to the crystal axes a, b, and c, and the results have to be summarized to an averaged absorption cross section $C_{abs}$. In the case of corundum and quartz, the polarization parallel to a and b results in identical spectra, which means that the absorption cross section for the ordinary ray ($C_{abs\,o}$) has to be taken with double weight 
compared to the absorption cross section for the extraordinary 
ray $C_{abs\,eo}$:

\begin{equation}
C_{abs}=\frac{1}{3}C_{abs\,eo}+\frac{2}{3}C_{abs\,o}.
\label{eq:Cabsav}
\end{equation}

In the case of spinel, the averaging is of course not necessary. It has 
to be noted
that for all nonspherical particle models, the separate treatment of the 
crystal axes
in calculating the absorption cross sections is physically not exact 
and can lead
to deviations in the predicted spectra from reality. This can 
only be
circumvented by much more complex models such as the discrete dipole 
approximation.
In the main bands of corundum and spinel, however, the deviations are 
minor, and
the spectra shown in the following are sufficiently reliable. For 
details see Mutschke et al.\ (\cite{Mutschke09}).

\begin{figure}
	\centering
	\includegraphics[width=\linewidth]{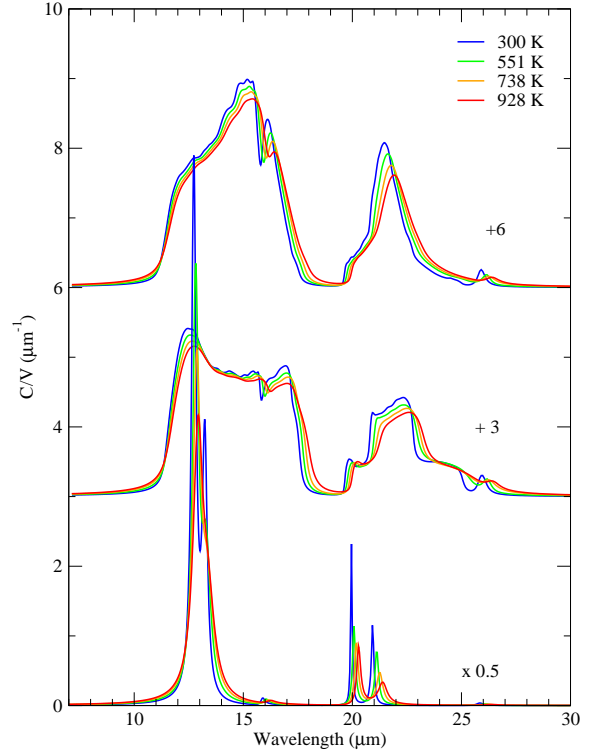}
	\caption{Comparison of the volume normalized $C_{abs}$ of small particles of corundum for Mie theory (multiplied by 0.5), a DFF model for aggregated spheres (offset +3), and a DFF model for Gaussian random spheres (offset +6) at different temperatures.}
	\label{fig:cor_all_1}
\end{figure}

\begin{figure}
	\centering
	\includegraphics[width=\linewidth]{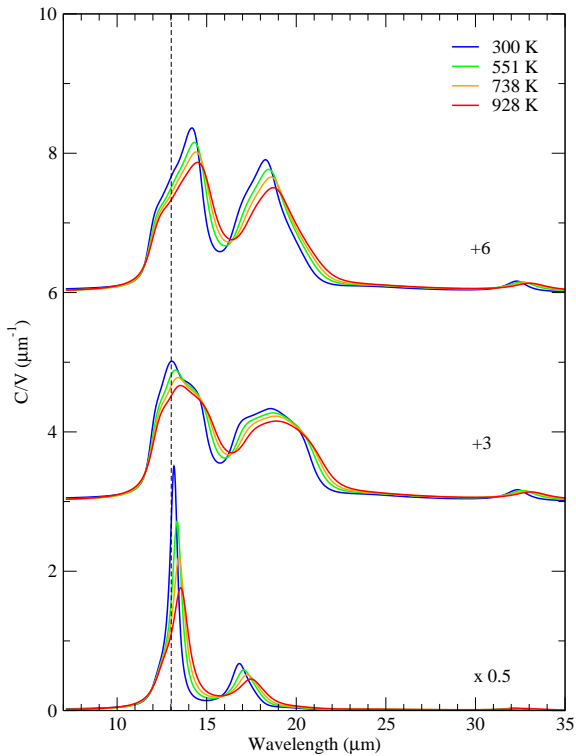}
	\caption{Comparison of the volume normalized $C_{abs}$ of small particles of spinel for Mie theory (multiplied by 0.5), a DFF model for aggregated spheres (offset +3), and a DFF model for Gaussian random spheres (offset +6) at different temperatures. The dashed vertical line indicates the 13\,$\mu$m band position.}
	\label{fig:spin_all}
\end{figure}

\begin{figure}
	\centering
	\includegraphics[width=\linewidth]{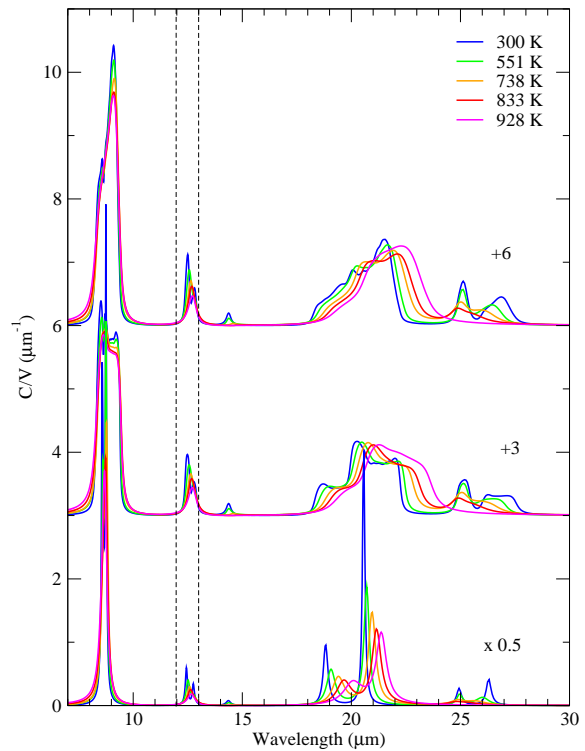}
	\caption{Comparison of the volume normalized $C_{abs}$ of small particles of quartz for Mie theory (multiplied by 0.5), a DFF model for aggregated spheres (offset +3), and a DFF model for Gaussian random spheres (offset +6) at different temperatures. The features that could account for the astronomical 13\,$\mu$m band are actually located between 12 and 13\,$\mu$m (hence the dashed vertical lines).}
	\label{fig:qu_all_lin}
\end{figure}

In Figs.\ \ref{fig:cor_all_1}--\ref{fig:qu_all_lin}, the calculated
{temperature-dependent} $C_{abs}$-spectra of corundum, spinel, and quartz are presented. In each figure, the spectra for small spheres (Mie theory), aggregates of spheres and Gaussian random spheres are compared. For all three minerals, the differences between the spectra of Mie theory (bottom spectra) and those of the DFF model (offset spectra) are the most prominent. In Mie theory, the strongest modes appear as very sharp and intense bands in the spectra of $C_{abs}$ while already for aggregated spheres (offset +3), the bands become much broader toward longer wavelengths, although a maximum remains at the position of the Mie peak. In the spectra of the Gaussian random spheres (offset +6), even the maximum of the bands becomes shifted toward longer wavelength, while only a shoulder remains at the position of the Mie peak. Therefore, the FWHM of the bands in the spectra of Gaussian random spheres is not very different to the one of the aggregates of spheres. Note that most changes in the $C_{abs}$-spectra take place only for the strongest bands. The weaker bands do not show any strong {particle shape dependence}.

\section{Consequences of our measurements for the astronomical 13\,$\mu$m band 
\label{sec:astro}}

The small-particle spectra derived in the previous sections
enable a better understanding of solid state bands, especially those detected 
in warm and hot dust-forming astronomical environments. Indeed, the most straightforward application of our laboratory spectra is the onset phase of dust-formation in AGB stars, where mainly minerals belonging to the hot end of the oxygen-rich condensation sequence form. This is because our data refer to
refractory minerals, and our main innovation is the systematic study of the high temperature spectra of these minerals.

Interestingly, there is a subset of dust-forming stars on the Asymptotic Giant Branch (AGB), where refractory dust species, such as corundum and/or spinel
seem to be over-represented compared to the silicates. (The latter tend to be the overwhelmingly dominating kind of dust for other oxygen-rich stars on the AGB.)
This subset is characterized by optically thin shells, in most cases by
a semiregular pulsational behavior (Sloan et al.\ 1996), and -- in
terms of IR spectroscopy -- by the presence of the so-called 13
micron feature,\footnote{About 90\% of the semiregular variables for
which IR spectra are available show the 13\,$\mu$m feature according
to Sloan et al.\ (1996).} which has been the subject of many papers (e.g.\
Glaccum 1995, Sloan et al.\ 1996, Posch et al.\ 1999, Fabian et al.\ 2001,
Sloan et al.\ 2003, DePew et al.\ 2006), and even of a PhD thesis (DePew
2006).

Consensus has been reached that this spectral feature is due to an oxide rather than to a silicate dust species. In fact, {\em all three} dust species discussed in this paper have been proposed as carrier of the 13\,$\mu$m band. 
Speck et al.\ (2000) propose (amorphous) SiO$_2$ as feature carrier, 
Posch et al.\ (1999) and Fabian et al.\ (2001) favor MgAl$_2$O$_4$,
Sloan et al.\ (2003) and DePew et al.\ (2006) have strengthened the case for 
$\alpha$-Al$_2$O$_3$. It should be noted that this dust feature has also been detected in AGB stars outside our galaxy (e.g.\ Lebzelter et al.\ 2006) 
and in the spectra of S-type AGB stars (Smolders et al.\ 2012).

As for SiO$_2$, the optical constants that we measured at high temperatures
do not make it a more viable candidate carrier of the 13\,$\mu$m band than the
room temperature data did. The by far strongest band of hot 
$\alpha$-SiO$_2$ is located at wavelength of 8.7--8.9\,$\mu$m for 
spherical grains and also remains around 9\,$\mu$m for the nonspherical
grain shapes we examined. However, no significant narrow 8.7--9 \,$\mu$m 
emission feature is detected in those astronomical objects that show the
13\,$\mu$m band. The sole (weak) resonance mode of hot $\alpha$-SiO$_2$ which
comes close to 13\,$\mu$m is in fact located around 12.5-12.7\,$\mu$m, with
only a weak dependence on the grain shape, as mentioned before
and as demonstrated in Fig.\ \ref{fig:qu_all_lin}.
As a result, $\alpha$-SiO$_2$ cannot be considered as a promising candidate
carrier of the 13\,$\mu$m band; so we focus on the application 
of our high temperature data for hot spinel and corundum.

The following procedure has been applied to compare the small-particle spectra derived from our optical constants with astronomical spectra:
i) each $C_{abs}$ spectrum (for our measured T$_d$ values) is 
multiplied with the corresponding black body function $B_\nu(T_{d})$, 
and the result is normalized to unity. We refer to the resulting curves, 
i.e.\ to the normalized $C_{abs} \times B_\nu(T)$ curves, as normalized dust emissivities.\\
ii) Selected spectra of AGB stars showing the 13\,$\mu$m bands are
treated as outlined by Posch et al.\ (1999) and Fabian et al.\
(2001) to derive residual dust emissions.
A polynomial is used to represent all other dust
components but the carrier of the 13\,$\mu$m band.
The respective polynomial is then subtracted from the spectra.
The result is also normalized to unity.
A mean out of 23 such residual dust emissions (based on ISO spectra)
was calculated.
In short, the following comparisons confront normalized dust emissivities
with mean residual dust emissions in the 11-15\,$\mu$m range.
Just for comparison, the residual emission spectrum of an S-type
AGB star as seen with {\em Spitzer}\/ -- taken from Smolders et al.\
(2012) -- is also shown.


\subsection{Hot corundum and the 13\,$\mu$m band}

As mentioned before, crystalline $\alpha$-Al$_2$O$_3$
has been considered as a potential carrier of the 13\,$\mu$m feature;
e.g., Glaccum (1995) wrote that ``hot sapphire is the most likely source
of the 13\,$\mu$m feature found in some M and MS stars'' (but he did not
indicate on which IR data this assertion was based). As for the term ``sapphire'', we will avoid it in the following, since it refers to
$\alpha$-Al$_2$O$_3$ containing coloring impurities, e.g.\ small
amounts of Fe$^{2+}$, Fe$^{3+}$, Ti$^{3+}$ or V$^{4+}$. We instead
refer to pure $\alpha$-Al$_2$O$_3$ and use the term corundum for it.
Evidence of the formation of Al$_2$O$_3$ came from extensive research on
presolar grains in meteorites (e.g.\ Nittler 1997, Clayton \& Nittler
2004; Hoppe 2004). However, it has so far not been possible to set any strong
constrain on the polytype of Al$_2$O$_3$  by presolar grain studies 
(which is a problem, since many polytypes of Al$_2$O$_3$ exist -- see,
e.g., Tamanai et al.\ 2009).

Until very recently, it has been impossible to produce any satisfying fit of the 
13\,$\mu$m band using optical constants of corundum at room temperature. As shown by Posch et al.\ (1999), the following dilemma arose, based on the room 
temperature data:\\
a) for small spherical particles, corundum would produce a 12.7\,$\mu$m feature
with too narrow a bandwidth;\\
b) for a continuous distribution of ellipsoids, $\alpha$-Al$_2$O$_3$ would
produce a 13.2\,$\mu$m feature with an up to 7 times too large bandwidth.

\begin{figure}
\centering
\includegraphics[width=1.0\linewidth]{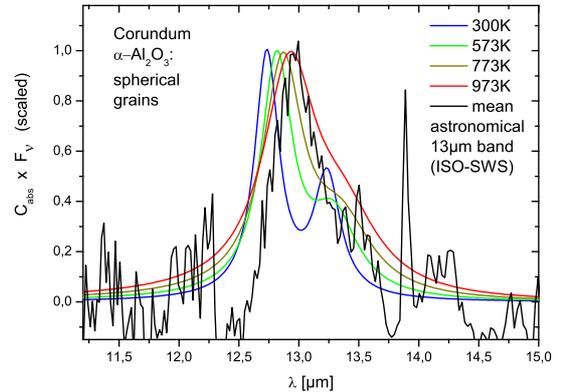}
\caption{An average profile of the astronomical
13$\mu$m emission band -- as derived by Fabian et al.\ (2001) for a
sample of 23 oxygen-rich AGB stars -- compared to the normalized emission
spectra of spherical corundum grains at temperatures of 300\,K
to 928\,K.}
\label{fig:cor_13mic}
\end{figure}

\begin{figure}
\centering
\includegraphics[width=1.0\linewidth]{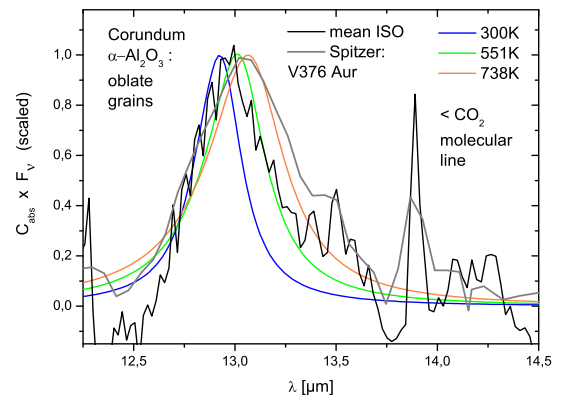}
\caption{The case of nonspherical corundum grains. These grains, particularly rotational ellipsoids with an axis ratio of r$_c$/r$_a$=0.79, enable a good representation of the observed 13\,$\mu$m band profile (the latter is the same as in the previous figure). In addition to our mean ISO 13\,$\mu$m band profile, a single Spitzer-IRS 13\,$\mu$m band profile (of V376 Aur, kindly provided by K.\ Smolders) is also shown.}
\label{fig:cor_13mic2}
\end{figure}

Takigawa et al.\ (2012) proposed to solving this dilemma by
introducing ellipsoidal grains of a particular shape. In their condensation
experiments, they found that the condensation rates of crystalline
Al$_2$O$_3$ are quite different for the different crystallographic axes,
which leads to a flattening of the condensed particles along the
crystallographic c-axis. They derive an axis ratio of r$_c$/r$_a$=0.79
and note that for oblate spheroidal grains with a slightly lower ratio
of r$_c$/r$_a$=0.70, the position of the 13\,$\mu$m feature could be well
reproduced with room temperature optical constants.
This is because, compared to spherical grains, such oblate spheroids have a resonance for the E$\perp$c contribution at slightly longer wavelengths, while the minor resonance of the E$\parallel$c contribution shifts to shorter
wavelengths. This leads to a single peak, which in the case of the
experimentally derived axis ratio r$_c$/r$_a$=0.79 would still be positioned a bit shortward of the 13\,$\mu$m feature and even narrower than the spherical
resonance. Allowing for a range of axis ratios around a somewhat changed
value of r$_c$/r$_a$=0.70 would give the right feature (cf.\ Takigawa et al.\ 
2012; see also Bohren \& Huffman 1983, Sect.\ 5.3, for the applied
method of calculating C$_{abs}$ for ellipsoidal grains).
We show in the Figs.\ \ref{fig:cor_13mic} and \ref{fig:cor_13mic2} 
how our high-temperature data partly confirm and partly modify these 
results.

Figure \ref{fig:cor_13mic} refers only to spherical grains. The plot shows
how the increasing temperature shifts the peak position of the strongest
resonance of spherical corundum grains to longer wavelengths, until a position
of almost exactly 13\,$\mu$m is reached for temperatures close to 1000\,K.
However, the bandwidth significantly increases with temperature. At the same
time, the peak height C$_{abs, max}$ is reduced, which is not seen in the
normalized spectra. It will be noticed that for T = 928\,K, the width
of the 13\,$\mu$m corundum band is too large compared to its (average) astronomical counterpart.
It should be noted, though, that there are individual astronomical
objects with slightly broader 13\,$\mu$m bands than others
(cf.\ Posch et al.\ 1999, Tab.\ 1 and Fabian et al.\ 2001,
Tab.\ 4). We also verified this with Spitzer-IRS spectra, kindly provided by 
K.\ Smolders. Broader band profiles may indicate higher (mean) dust 
temperatures.

Figure \ref{fig:cor_13mic2} refers to oblate spheroidal grains 
with the experimental axis ratio found by Takigawa et al.\ (2012),
i.e.\ r$_c$/r$_a$=0.79. With the room temperature data, we find, in agreement with their result, that the calculated band is at a somewhat shorter wavelength and that a further flattening of the particle shape would be necessary for a perfect match.
For grains radiating at T=551\,K, however, a very good 
representation of the observed 13\,$\mu$m band profile is achieved.
For still higher temperatures, the corundum peak is shifted 
even further to the red (see the curve for T=738\,K in Fig.\ 
\ref{fig:cor_13mic2}). This leads to a discrepancy with our mean
(ISO-SWS-based) profile of the 13\,$\mu$m band (see below).

It should be noted that only {\em one percent}\/ of corundum in the
circumstellar shell of an AGB star are sufficient to account for the strength 
of the 13\,$\mu$m band. This has been shown by DePew et al.\ (2006). In this study, the discrepancy between the peak position of corundum spheres and the observed 13\,$\mu$m band had not become evident due to a relatively 
coarse wavelength grid (used to speed up the radiative transfer calculations, typically with one grid point every 0.5\,$\mu$m around 13\,$\mu$m; see their 
Fig.\ 5).

The study by DePew et al.\ (2006) also reveals how it may be possible that we do see corundum's (or spinel's) $\sim$13\,$\mu$m feature, while we {\em do not see their weaker bands at longer wavelengths:} those bands may simply be suppressed by temperature effects and owing to the superposition of other dust components
(see also the final section of Sloan et al.\ 2003).

Another argument (other than Fig.\ \ref{fig:cor_13mic2}) for corundum being the carrier of the 13\,$\mu$m band fact that this band is frequently detected together with a broad spectral feature peaking at $\sim$11--12\,$\mu$m,
which is usually assigned to amorphous Al$_2$O$_3$. The coexistence of crystalline corundum 
and amorphous Al$_2$O$_3$ in an oxygen-rich circumstellar shell with a low mass loss rate seems quite likely. Incomplete condensation of a gas of solar composition
might at the same time result in the formation of amorphous alumina
and in a relative lack of silicates (compared to shells with greater optical depths that had more time to form silicates as well, in addition to refractory oxides -- see, e.g., Smolders et al.\ 2012 for a short discussion of this idea).


\subsection{Hot spinel and the 13\,$\mu$m band}

For the case of spinel, we restrict ourselves on the small 
particle spectra of {\em spherical}\/ grains -- since, as pointed out by 
DePew et al.\ (2006), nonspherical grains of spinel can hardly account for the specific properties (bandwidth, bandshape) of the 13\,$\mu$m band. More specifically, a distribution of ellipsoidal grains or hollow spheres
leads to a broadening and shift to longer wavelengths of spinel's main MIR emission band, which is incompatible with the properties of the 13\,$\mu$m
feature (see also Posch et al.\ 1999).

\begin{figure}
\centering
\includegraphics[width=1.0\linewidth]{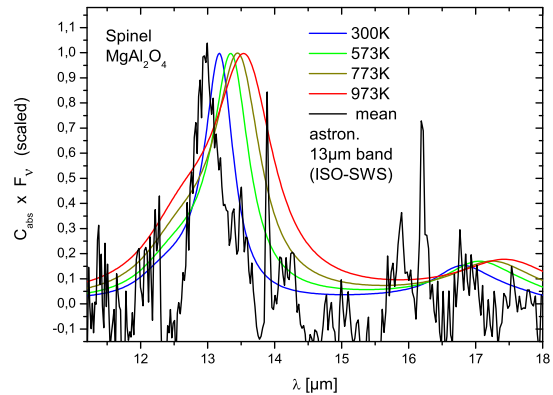}
\caption{Comparison between the average profile of the astronomical
13$\mu$m emission band -- same as in the previous figure -- with the normalized emission spectra of spherical spinel grains at temperatures of 300\,K 
to 928\,K.}
\label{fig:spin_13mic}
\end{figure}

In Fig.\ \ref{fig:spin_13mic}, the same astronomical spectra as in the previous
figures are shown in comparison to the high temperature emissivities of spinel grains. It becomes evident from this figure that the 300\,K emissivity of spinel
-- i.e.\ its room temperature spectrum! -- comes closest to the profile of the
13\,$\mu$m band. For all the higher temperatures, the bandwith of spinel's strongest emission signature becomes much broader than for the astronomical counterpart (with the FWHM reaching 1.38\,$\mu$m at 928\,K). The peak position
shifts to 13.5\,$\mu$m for the largest examined temperature 928\,K, which is also at odds with the peak wavelength of the 13\,$\mu$m feature.
For nonspherical spinel grains, the 13\,$\mu$m band peak position shifts to even slightly longer wavelengths than 13.5\,$\mu$m 
(see Fig.\ \ref{fig:spin_all}).
Table \ref{t:summary} summarizes essential properties of the observed
13\,$\mu$m band and of its potential carriers (spinel {\em and} corundum)
radiating at different temperatures.

\begin{table}[htbp]
\caption{Comparison of the properties of the observed 13\,$\mu$m band 
profile with those of spinel and corundum at different temperatures,
according to our new in situ high temperature optical constants.}
\label{t:summary}
\begin{center}
\begin{tabular}{llll}
\hline
substance    & $\lambda_{\rm peak}$  & FWHM     & comments \\
or spectrum  & [$\mu$m]              & [$\mu$m] &          \\
\hline
Mean ISO-SWS             & 13.00  & 0.45  &  \\
V376 Aur, Spitzer-IRS      & 13.07  & 0.60  &  \\
corundum, spher., 300\,K   & 12.73  & 0.26 & + shoulder  \\
                           &       &      & at 13.2\,$\mu$m  \\
corundum, spher., 551\,K   & 12.82  & 0.35 &  \\ 
corundum, spher., 738\,K   & 12.87  & 0.50 &  \\
corundum, spher., 928\,K   & 12.94  & 0.73 &  \\
corundum, obl., 300\,K   & 12.92  & 0.27 &  \\
corundum, obl., 551\,K   & 13.00  & 0.37 &  \\ 
corundum, obl., 738\,K   & 13.07  & 0.46 &  \\
spinel, spher., 300\,K   & 13.17  & 0.53 &  \\
spinel, spher., 551\,K   & 13.35  & 0.72 &  \\ 
spinel, spher., 738\,K   & 13.44  & 0.97 &  \\
spinel, spher., 928\,K   & 13.53  & 1.37 &  \\
\hline
\end{tabular}
\end{center}
\end{table}

The peak in the mean ISO spectrum at 16.8\,$\mu$m seems to coincide with a
minor emissivity peak of spinel at room temperature, as noted by Posch 
et al.\ (1999) and Fabian et al.\ (2001). However, the 16.8\,$\mu$m
band seen in some astronomical sources of the 13\,$\mu$m feature has meanwhile
been assigned to a CO$_2$ molecular line (Sloan et al.\ 2003). Therefore, this small spectral band can no longer be used to support the assignment of the 13\,$\mu$m band to spinel.

\begin{figure}
\centering
\includegraphics[width=1.0\linewidth]{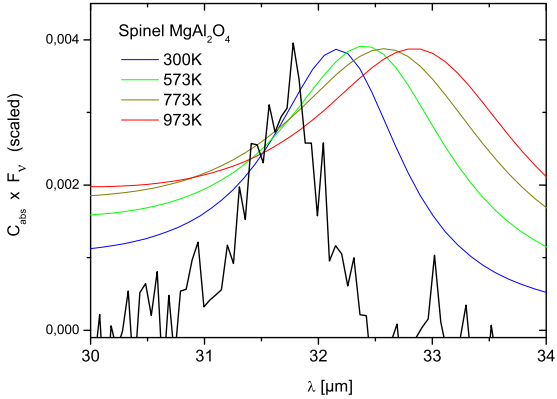}
\caption{Comparison between the average profile of the astronomical
32$\mu$m emission band with the normalized emission spectra of spherical 
spinel grains at temperatures of 300\,K to 928\,K.}
\label{fig:spin_32mic}
\end{figure}

There is yet another spectral feature that spinel grains, if present in a
circumstellar shell, are supposed to produce, namely an emission band peaking close to 32\,$\mu$m (Fabian et al.\ 2001).
Indeed, there is a narrow emission feature at 31.8\,$\mu$m in many of the sources of the 13\,$\mu$m band.
Therefore, we show an average profile of this 31.8\,$\mu$m band (according to
Posch et al.\ 2006) and compare it to the emissivity of hot spinel grains
in Fig.\ \ref{fig:spin_32mic}. This comparison shows that even for
spherical grains at room temperature, no satisfying fit of the astronomical 
31.8\,$\mu$m band with stoichiometric spinel can be achieved (the bandwidth
for spinel at T=300\,K is too large, and the peak position is at too large
a wavelength).
For spinel at higher temperatures, our new optical constants lead to an even
larger discrepancy between its emissivity and 31.8\,$\mu$m band profile.
We rather expect that spinel grains {\em cooler}\/ than 300\,K produce a
31.8\,$\mu$m band with the required (narrow) bandwidth, but presently 
available data do not yet allow a decision on this question.

\section{Concluding remarks}

Our new optical constants show that both spinel and corundum grains, if  radiating at a suitable narrow range of inermediate temperatures, can 
account for the average profile of the astronomical 13\,$\mu$m band: 
spinel at about 300--400\,K and corundum at about 500--700\,K 
(also depending on the grain shapes).

At still higher grain temperatures, the resulting integrated emission,
both of spinel and of corundum grains, becomes too broad as to fit
the observed astronomical band. The precise temperature at which
spinel's and/or corundum's 13\,$\mu$m band becomes too broad and too
strongly `redshifted' compared to the observations depends on the individual astronomical object under examination, since some observed 13\,$\mu$m bands
of AGB stars are broader and/or peak at longer wavelengths than others.
Therefore, the 13\,$\mu$m band may even be used as a
dust temperature indicator, at least at a somewhat more advanced stage of
astronomical MIR spectrosopy (with even more available MIR spectra of AGB stars
and with increased signal-to-noise ratios).

So far, we have fitted only the 13\,$\mu$m band with our new optical constants of hot corundum and spinel. Both of them have additional
weaker bands at longer wavelengths for which no counterparts in
the observed astronomical spectra could be found so far. For spinel's
32\,$\mu$m feature, there is an astronomical counterpart, but at a smaller
peak wavelength and with a narrower bandwidth (only compatible with
temperatures of MgAl$_2$O$_4$ grains below 300\,K).

The lack of (some) longer wavelength features may be due to a suppression
of those bands by temperature
effects, as shown by other authors (e.g.\ de Pew et al.\ 2006).
Radiative transfer calculations should be made again, based on our
optical constants,
in order to precisely predict the strengths of those secondary bands
(at an expected fractional abundance of corundum and/or spinel amounting
to a few percentage points) and to check systematically with which fitting
parameters such predictions are indeed compatible with presently available
astronomical spectra.


\begin{acknowledgements}

We thank Gabriele Born, Jena, for the sample preparation and for help with the EDX measurements. Kristof Smolders, Leuven, kindly provided Spitzer spectra of 
V376 Aur. Furthermore, we are very thankful to an anonymous referee who helped us improve the paper. This project is a part of the SPP (\textit{``Schwerpunktprogramm}'') of the DFG (Deutsche Forschungsgemeinschaft):
\textit{``The First 10 Million Years of the Solar System''}.

\end{acknowledgements}


\end{document}